\pdfoutput=1
\documentclass[aps, twocolumn, showpacs, superscriptaddress, prl]{revtex4-1}

\usepackage[pdftex]{graphicx}
\usepackage[pdftex]{epsfig}
\usepackage{epstopdf}
\usepackage{color}
\usepackage{inputenc}
\usepackage{amssymb,amsmath}
\usepackage[hypertexnames=false,colorlinks,urlcolor=blue,citecolor=blue]{hyperref}
\usepackage{braket}
\usepackage{siunitx}
\usepackage{multirow}
\usepackage{mathtools}
\usepackage{enumitem}
\usepackage{qcircuit}

\def\block(#1,#2)#3{\multicolumn{#2}{c}{\multirow{#1}{*}{$ #3 $}}}

\usepackage[dvipsnames]{xcolor}
\definecolor{gosharks}{rgb}{0, 0.42, 0.45}
\newcommand{\comment}[1]{}

\newcommand{\stefan}[1]{}
\newcommand{\aer}[1]{}

\newcommand{\ags}{\Phi_0}
\newcommand{\gs}{\Psi_0}
\newcommand{\agsi}{\Phi_{0,i}}
\newcommand{\agsf}{\Phi_{0,f}}

\newcommand{\hamiltonian}{\mathcal{H}}
\newcommand{\overlap}{\gamma}  
\newcommand{\improvement}{\iota}

\newcommand{\prep}[1]{{\rm PREP}_{#1}}
\newcommand{\prept}[1]{\widetilde{{\rm PREP}}_{#1}}
\newcommand{\unprep}[1]{{\rm UNPREP}_{#1}}
\newcommand{\sel}[1]{{\rm SEL}_{#1}}
\newcommand{\norm}[1]{\alpha_{#1}} 

\newcommand{\ubmltext}[1]{\text{\scriptsize\tabular[t]{@{}c@{}}#1\endtabular}}

\newcommand{\papertitle}{Quantifying $T$-gate-count improvements for ground-state-energy estimation with near-optimal state preparation}

\begin{document}

\title{\papertitle}

\author{S. Pathak}
\affiliation{Quantum Algorithms and Applications Collaboratory, Sandia National Laboratories, Albuquerque NM, USA}
\author{A.E. Russo}
\affiliation{Quantum Algorithms and Applications Collaboratory, Sandia National Laboratories, Albuquerque NM, USA}
\author{S.K. Seritan}
\affiliation{Quantum Algorithms and Applications Collaboratory, Sandia National Laboratories, Livermore CA, USA}
\author{A.D. Baczewski}
\affiliation{Quantum Algorithms and Applications Collaboratory, Sandia National Laboratories, Albuquerque NM, USA}

\begin{abstract}
We study the question of when investing additional quantum resources in preparing a ground state will improve the aggregate runtime associated with estimating its energy.
We analyze Lin and Tong's near-optimal state preparation algorithm and show that it can reduce a proxy for the runtime, the $T$-gate count, of ground state energy estimation near quadratically.
Resource estimates are provided that specify the conditions under which the added cost of state preparation is worthwhile.
\end{abstract}

\maketitle

\textit{Introduction.---}
A key task in all quantum simulation algorithms is the preparation of a state that encodes the observables of a physical system of interest~\cite{lloyd1996universal,aspuru2005simulated,jordan2012quantum}.
This is often the ground state $\ket{\gs}$ of a Hamiltonian $\hamiltonian$ on an $n$-qubit Hilbert space~\cite{poulin2009preparing,tubman2018postponing,ge2019faster,lin2020near,lemieux2021resource}.
Independent of whether this represents interacting electrons~\cite{reiher2017elucidating,babbush2018encoding,lee2021even,su2021fault,su2021nearly,von2021quantum}, spins~\cite{childs2018toward,childs2019nearly,tran2021faster}, or quantum fields~\cite{klco2019digitization,lamm2019general,shaw2020quantum}, we generally produce an approximation to the desired state $|\ags\rangle$ with overlap $\overlap=|\langle \ags|\gs\rangle|$.
Then the probability of successfully processing $\ket{\ags}$ to estimate some observable of interest (e.g., the ground state energy $E_0$) is generally upper bounded by $\overlap^2$~\cite{nielsen2002quantum}, which would ideally be 1.

While it is likely not possible to efficiently prepare ground states of \emph{generic} local Hamiltonians on quantum computers~\cite{kitaev2002classical,kempe2006complexity} physical arguments suggest that the \emph{specific} instances for which nature can efficiently find the ground state will be efficiently preparable on a quantum computer~\cite{feynman1982simulating}.
Though the question of which instances these are remains an active area of research~\cite{ogorman2022intractability}, it is generally of interest to develop algorithms that increase $\overlap_i$ from some easy-to-prepare initial approximation, $\ket{\agsi}$, to $\overlap_f$ with some associated final approximation $\ket{\agsf}$.
It might be that $\ket{\agsi}$ comes from the outcome of a classical calculation (e.g., an approximate solution to a mean-field theory) or a hybrid quantum-classical approach like the variational quantum eigensolver~\cite{peruzzo2014variational,omalley2016scalable}. 
Regardless of the source of $\ket{\agsi}$, it is often assumed that the cost of preparing it is negligible relative to the cost of boosting $\overlap_i \rightarrow \overlap_f$, making use of some unitary $\mathcal{U}_{SP}(\hamiltonian)\ket{\agsi}=\ket{\agsf}$~\footnote{This cost isn't always negligible. As has been noted in Ref.~\cite{delgado2022Apr}, sometimes even the cost of preparing simple input states can dominate simulation costs unless certain simplifying assumptions are made.}.

\begin{figure*}[ht]
\includegraphics[width=\textwidth]{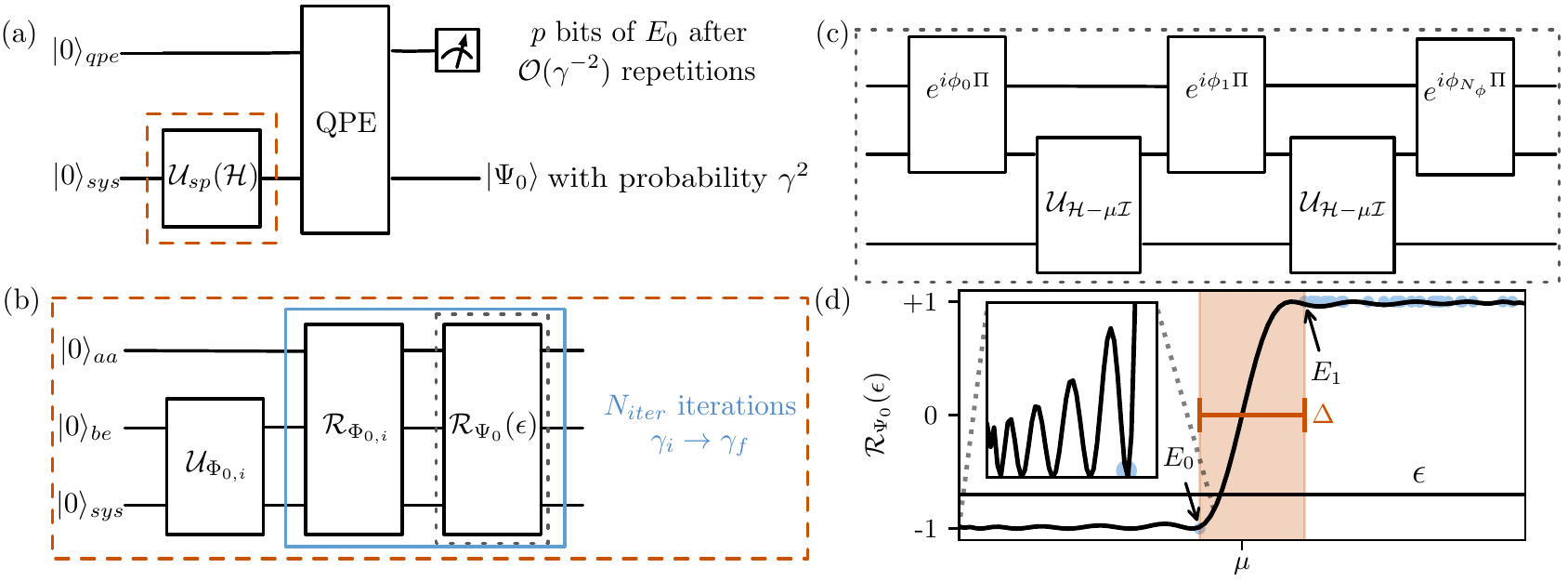}
\caption{Circuit diagrams and schematic visualizations for estimating the ground state energy $E_0$ of a Hamiltonian $\hamiltonian$ with a near-optimal ground state preparation technique using amplitude amplification.
(a) An energy estimation circuit like the one in the reference implementation under consideration~\cite{babbush2018encoding} in which the probability of correctly measuring $E_0$ in a single round is $\overlap^{2}$. 
The value of $\overlap$ is determined by details of $\mathcal{U}_{sp}(\hamiltonian)$.
(b) A circuit that implements $\mathcal{U}_{sp}$ by first preparing an initial guess $\ket{\agsi}$ and then running $N_{iter}$ rounds of AA in which the reflector about $\ket{\agsi}$ ($\mathcal{R}_{\agsi}$) is straightforward to implement and the reflector about $\ket{\gs}$ ($\mathcal{R}_{\gs}(\epsilon))$ requires the use of quantum signal processing, as proposed in Ref.~\cite{lin2020near}, to produce $\ket{\agsf}$.
$N_{iter}$ is determined by $\overlap_i$ and the desired final overlap $\overlap_f$.
The former overlap might vanish exponentially in the size of $sys$.
(c) A circuit that uses quantum signal processing to implement an $\epsilon$-approximation to a degree-$N_\phi$ polynomial that suppresses support on energy eigenstates above a known lower bound ($\mu$), in favor of support on energy eigenstates below $\mu$. 
The phases are chosen using the method in Ref.~\cite{dong2021efficient}.
(d) An illustration of the $\epsilon$-approximate reflector, $\mathcal{R}_{\Psi_0}(\epsilon)$, and how it acts on eigenstates above and below the energy gap $\Delta$.
Here, the calculated phases realize an approximation that clearly outperforms the target $\epsilon=0.3$.
\label{fig:circuit_diagram}
}
\end{figure*}

The question that we answer in this Letter is ``when does the added cost of implementing $\mathcal{U}_{SP}(\hamiltonian)$ outweigh the cost of repeated trials with a lower probability of success?''
We are specifically concerned with the potential benefit to estimating $E_0$, for which conventional approaches that apply quantum phase estimation (QPE)~\cite{kitaev1995quantum} to $\ket{\agsi}$ will project onto $\ket{\gs}$ after a single round with probability $\overlap_i^2$~\cite{nielsen2002quantum} (see Fig.~\ref{fig:circuit_diagram}(a)).
One can use more elaborate strategies in which repeated rounds of QPE can iteratively improve our knowledge of $E_0$~\cite{aspuru2005simulated,kimmel2015robust,wiebe2016efficient,o2021error,russo2021evaluating,lin2022heisenberg,wang2022quantum}, though we assume a simple strategy of repeating the circuit $\mathcal{O}(\overlap_i^{-2})$ times for ease of analysis~\footnote{We expect that further improvements can be made by adopting the aforementioned ``more elaborate strategies'' but these are likely to be constant-factor improvements rather than asymptotic ones.}.

Broadly speaking there are two classes of state preparation algorithms, those that make use of the adiabatic theorem~\cite{farhi2000quantum} and those that apply a filter in the eigenbasis of $\hamiltonian$~\cite{poulin2009preparing}.
In this Letter, we consider a $\mathcal{U}_{SP}(\hamiltonian)$ in the latter category, derived from a near-optimal approach of Lin and Tong~\cite{lin2020near}.
While adiabatic state preparation is conceptually straightforward, it generally requires time-dependent Hamiltonian simulation, the analysis of a family of Hamiltonians along the entire adiabatic pathway, and potentially many different initial Hamiltonians, rather than a single instance.
It also has worse scaling with the minimum spectral gap~\cite{albash2018adiabatic}.
However, there are some important exceptions~\cite{wan2020fast} and it may be the case that adiabatic algorithms are found to be the optimal solution in some cases, so we make no claim to the optimality of our results.
An added benefit of analyzing filter-based state preparation is that it relies on a block encoding~\cite{gilyen2019quantum} of $\hamiltonian$ that could be identical to one used in QPE, making it straightforward to compare costs.

Whether exponential, polynomial, or nonexistent, the advantages realized by quantum computers in physical simulation are likely to be problem-specific and depend critically on the cost of implementing $\mathcal{U}_{SP}(\hamiltonian)$~\cite{lee2022is}.
We provide estimates for the cost of state preparation based on the $T$-gate count of implementations of Lin and Tong's $\mathcal{U}_{SP}(\hamiltonian)$~\cite{lin2020near}. 
We choose the total number of $T$ gates as a simple-to-compute proxy for an actual runtime estimate because their implementation will dominate the runtime in T-factory-limited surface code architectures~\cite{fowler2012surface,litinski2019magic} and a more precise analysis would involve detailed scheduling of the algorithm's implementation.
Actual runtimes could also be reduced relative to this proxy in contexts in which the availability of magic states is not a limiting factor, in which case the $T$ depth would be the more appropriate quantifier.

We propose a ratio $\improvement$ that compares the $T$-gate count for $\mathcal{U}_{SP}(\hamiltonian)$ and subsequent QPE to the count for ``trivial'' state prep and QPE repeated until success.
This quantifies the improvement in runtime associated with ``better'' state preparation and allows us to answer the titular question.
Even in a scenario where $\overlap_i$ vanishes exponentially with increasing $n$, there are parameter regimes where there is a robust near-quadratic speedup for better state preparation.
This is consistent with the Grover-like speedup~\cite{grover1996fast} that one would expect~\cite{brassard2002quantum}, though we note that we are exploring this in terms of $T$-gate counts as a proxy for runtime, instead of query complexity or some other more abstract quantifier.

\textit{Methods.---}
Circuit diagrams that illustrate the implementation costs being studied are provided in Fig.~\ref{fig:circuit_diagram}.
The quantum computer that runs these circuits consists of four registers: ($sys$) the $n$-qubit system register that encodes $\ket{\agsi}$, ($qpe$) the auxiliary register  that  encodes $p$ bits of an estimate for $E_0$, ($aa$) which implement amplitude amplification (AA), and ($be$) which block encode $\hamiltonian$.
In what follows, we describe the structure of the circuits in Fig.~\ref{fig:circuit_diagram} and any attendant assumptions.
Many details that were originally elaborated elsewhere in the literature are summarized in the Supplemental Materials (SM)~\cite{SMref}.

For QPE, we use the highly optimized implementation of Babbush \textit{et al.}~\cite{babbush2018encoding}.
We indicate the $T$ count associated with estimating $E_0$ with Holevo variance $\Delta E$ as $T_{QPE}(\Delta E)$ and note that more relevant details can be found in the SM~\cite{SMref}.
We assume a particularly simple approach to estimating $E_0$.
Each repetition of the circuit will sample an eigenvalue from the spectrum of $\mathcal{H}$ with probability proportional to the overlap of the input state state with the associated eigenstate.
So after $\mathcal{O}(\overlap^{-2})$ repetitions the smallest observed eigenvalue will likely be $E_0$.
We now derive conditions under which boosting $\overlap^2$ with AA will reduce the expected total runtime for estimating $E_0$ relative to only relying on $\mathcal{U}_{\agsi}$, the cost of which we will denote $T_{\agsi}$ (see Fig.~\ref{fig:circuit_diagram}(b)).

AA consists of $N_{iter}$ applications of a product of two reflections ($\mathcal{R}_{\agsi}$, $\mathcal{R}_{\gs}$) that boosts the overlap of the state in $sys$ to $\overlap_f$.
$N_{iter}$ is determined by $\overlap_i$,
\begin{equation}
    N_{iter} =  \left\lceil \frac{1}{2}\left(\frac{\sin^{-1}\overlap_f}{\sin^{-1}\overlap_i} - 1\right)\right\rceil.
\label{eq:niter}
\end{equation}
While implementing $\mathcal{R}_{\agsi}$ only requires controlled applications of $\mathcal{U}_{\agsi}$, the implementation of $\mathcal{R}_{\gs}$ is complicated by $\ket{\gs}$ generally being unknown. 
We will follow Ref.~\cite{lin2020near} and construct an $\epsilon$-approximation to this reflector using quantum signal processing (QSP)~\cite{low2017optimal}, $\mathcal{R}_{\gs}(\epsilon)$ (see Fig.~\ref{fig:circuit_diagram}(c)).
This is a degree-$N_{\phi}$ polynomial in $(\hamiltonian-\mu \mathcal{I})$ that approximates a function that is ideally $-1$ for states with energy less than $\mu - \Delta/2$ and ideally $1$ for states with energy greater than $\mu+\Delta/2$.
Details pertaining to the calculation of $\mu$ and $\Delta$ are included in the SM~\cite{SMref}.

However, in practice each eigenvalue of $\mathcal{R}_{\gs}(\epsilon)$ is only $\epsilon$-close to $\lbrace -1,+1\rbrace$ and $\mathcal{R}_{\gs}(\epsilon)$ is not an exact reflector.
One of the technical advances in this Letter is a bound on $\epsilon$ such that the QSP circuit produces $\ket{\agsf}$ with $\overlap_{f} \geq \overlap_{i}$,
\begin{equation}
    \epsilon \leq \left(1 - \overlap_f^2\right)/6N_{iter}^2.
    \label{eq:approximation_bound}
\end{equation}
A proof can be found in the SM~\cite{SMref}.
The cost of implementing the entire AA circuit is denoted $T_{AA}(\overlap_i,\overlap_f)$.

The only remaining details for the implementation under consideration pertain to the block encoding of $\hamiltonian$~\cite{gilyen2019quantum}.
Block encoding is used both to encode the eigenspectrum of $\hamiltonian$, as sampled in QPE, and to implement a polynomial in $(\hamiltonian-\mu \mathcal{I})$ in QSP.
As many of the details are model-specific and extensively developed in other work, we relegate a detailed discussion of block encoding to the SM~\cite{SMref}.
While there are alternatives to block encoding, we leave it to future work to consider variants on the approach in Fig.~\ref{fig:circuit_diagram} making use of, e.g., Trotterized Hamiltonian evolution, either for encoding the eigenspectrum of $\hamiltonian$ for QPE or for implementing time-dependent Hamiltonian evolution in adiabatic state preparation~\cite{kocia2022digital}. 

With all of the components of our implementation specified, we can prepare $T$ counts for the circuits with and without AA.
The ratio of these counts defines the improvement,
\begin{equation}
  \improvement = \frac{\overlap_i^{-2}(T_{\agsi} + T_{QPE}(\Delta E))}{\overlap_f^{-2}(T_{\agsi} + T_{QPE}(\Delta E) + T_{AA}(\overlap_i, \overlap_f))}.
\label{eq:improvement_defn}
\end{equation}
Here the cost of $AA$ is
\begin{equation}
    T_{AA} = N_{iter}\Big(T_{\mathcal{R}_{\agsi}} + 
    N_{\phi}(T_{\mathcal{U}_{\hamiltonian}} + T_{e^{i\phi \Pi}})\Big).
    \label{eq:t_aa_cost}
 \end{equation}
$T_{\mathcal{R}_{\agsi}}$ involves two applications of $\mathcal{U}_{\agsi}$ and a single multi-controlled $X$ gate.
$T_{\mathcal{U}_{\hamiltonian}}$ is the cost of a single application of the block-encoded Hamiltonian.
$T_{e^{i\phi \Pi}}$ involves two multi-controlled $X$ gates and a single-qubit rotation with angles determined using the protocol in Ref.~\cite{dong2021efficient}.
$N_{\phi}$ is the number of phases used to implement $\mathcal{R}_{\gs}(\epsilon)$ and all parameter values will be chosen to saturate their bounds.
We note that Eq.~\ref{eq:t_aa_cost} depends on the rotation synthesis error incurred in implementing the Hamiltonian block encoding and the controlled rotations in QSP, and the error analysis used in deriving our results is considered in the SM~\cite{SMref}.
We consider better state preparation as being worthwhile when $\iota > 1$.

\textit{Results.---}
We first consider resource estimates for the 1D transverse field Ising model (TFIM)~\cite{tfimreview} with periodic boundary conditions~\footnote{Open boundary conditions will have very similar costs.}.
$sys$ is encoded such that each of the $n$ qubits represents one of the $L$ sites.
We consider a simple form for $\mathcal{U}_{\agsi}$ in which $R_y$ rotations are applied to each qubit to generate a product state.
We tune the rotation angles to construct a $\ket{\agsi}$ with a target value of $\overlap_i$.
While not a particularly sophisticated choice for initializing $sys$, it suffices for our purposes.
We select $\epsilon$ to saturate the bound in Eq.~\ref{eq:approximation_bound} with a target $\overlap_f^2$ = 0.75, and assume that the true final overlap is also $\overlap_f^2 $ = 0.75; a presentation of the difference between the target and true overlaps can be found in Fig.~\ref{fig:error_bound_results}.

\begin{figure}
\includegraphics{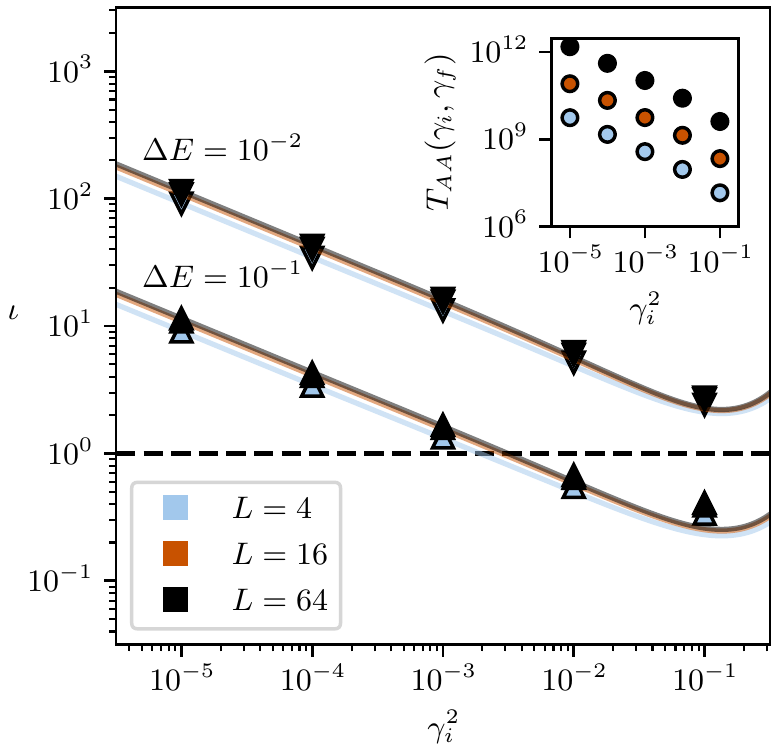}
\caption{$T$ counts and improvement $\iota$ for TFIM with $L = $ 4, 16, and 64 sites, with $\epsilon$ saturating the bound in Eq.~\ref{eq:approximation_bound} with $\overlap_f^2$ = 0.75, $\overlap_i^2$ ranging from $10^{-5}$ to $10^{-1}$, and two different values of $\Delta E$.
The inset shows $T_{AA}$ and the main figure is a plot of the improvement $\iota$ with solid lines fit to the asymptotic form Eq.~\ref{eq:asymptotic_improvement}.
The dashed line is a guide for the eye at $\iota = 1$, above which improvement is seen when conducting state preparation.
\label{fig:tfim_results}
}
\end{figure}

In Fig.~\ref{fig:tfim_results} we present $\iota$ and $T_{AA}$ for the TFIM as a function of $L$, $\overlap_i$, and $\Delta E$. 
We find $\iota > 1$ for all instances with $\overlap_i^2 \leq 10^{-3}$, and even for larger $\overlap_i$ for the higher accuracy calculation.
For $\overlap_i \ll 1$, the costs of QPE and AA are both dominated by repeated applications of the block-encoded Hamiltonian, leading to a simple approximate form of $\iota$
\begin{equation}
    \iota \sim \frac{\overlap_f^2}{\overlap_i^2} \Bigg(\frac{ \Delta}{\Delta E}\frac{1}{\sin^{-1}\overlap_f}\frac{\overlap_i}{\log\ \overlap_i^{-2}}\Bigg),
    \label{eq:asymptotic_improvement}
\end{equation}
consistent with a near-quadratic Grover-like speedup in $\overlap_i$ from AA.
Importantly, the asymptotic improvement has no explicit dependence on the system size, with the only system size dependence in the finite-size error of $\Delta$.
We find that our computed $\iota$ follows this asymptotic trend closely for $\iota \leq 10^{-2}$.
We note that 88 logical qubits are required for the $L = 64$ calculation with $\Delta E = 10^{-2}$, with a detailed analysis of qubit counts in the SM~\cite{SMref}.

To test the performance of our $\mathcal{U}_{sp}(\hamiltonian)$ implementation, we explicitly simulate circuits for $L = 2, 4, 6$ site TFIM using the upper bound on $\epsilon$ in Eq.~\ref{eq:approximation_bound}.
For each $L$ we run nine simulations, with $\overlap_f^2$ = 0.9, 0.99 and 0.999, and $\overlap_i$ chosen such that $N_{iter}$ = 4, 6, 10.
We are then able to compare the values of $\overlap_f^2$ actually realized in the simulation to the specified value of $\overlap_f^2$ in Fig.~\ref{fig:error_bound_results}.
Noise-free simulations were carried out using the PyTKET package~\cite{sivarajah2020pytket} with the full source available in the SM~\cite{SMref}.

\begin{figure}
\includegraphics{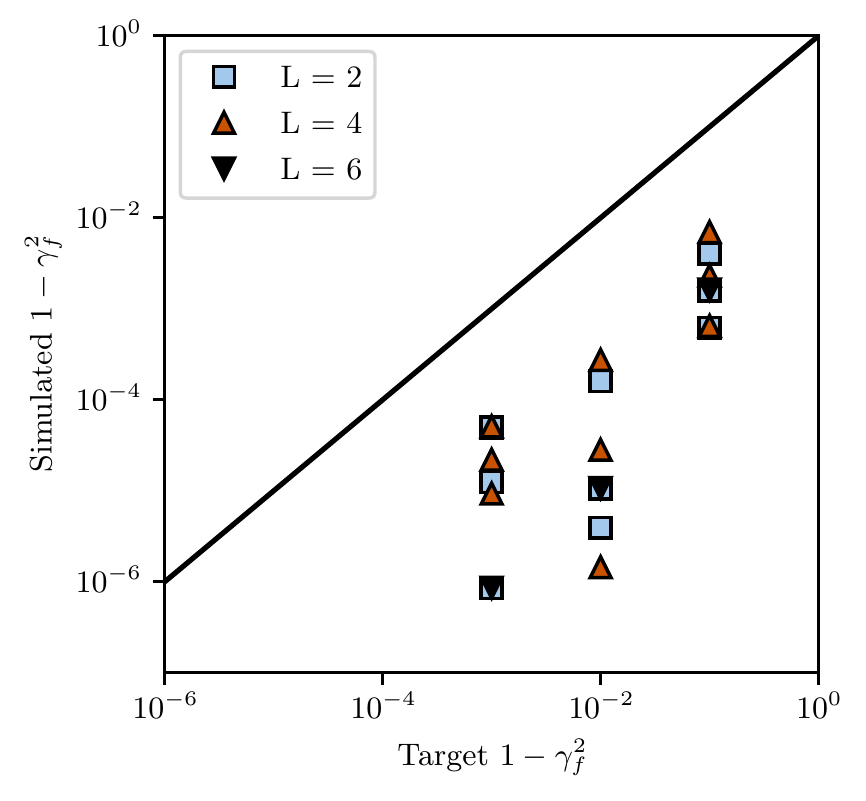}
\caption{Comparison of the target $\overlap_f$ and simulated $\overlap_f$ for the transverse field Ising model with $L=2$, $4$, and $6$ sites.
For each value of $L$, state-vector simulations of the full $\mathcal{U}_{sp}(\hamiltonian)$ circuit were carried out for three target values, $1 - \overlap_f^2 = 10^{-1}, 10^{-2}, 10^{-3}$, and three values of $\overlap_i$ satisfying $N_{iter} = 4, 6, 10$.
The black line is the $x = y$ reference and our results are consistent with the validity of the bound in Eq.~\ref{eq:approximation_bound}.
\label{fig:error_bound_results}}
\end{figure}

We find that the simulated infidelity, $1 - \overlap_f^2$, is consistently one or two orders of smaller than the target, indicating that our bound for $\epsilon$ in Eq.~\ref{eq:approximation_bound} could be adjusted to potentially realize further savings in implementing this approach to state preparation.
However, the impact of $\epsilon$ on $\iota$ is only logarithmic, so removing this looseness will only affect our results by a small constant factor.
As such, we are confident that our conclusions are not skewed by a loose bound in the state preparation $T$ counts.

Next, we consider resource estimates for a more realistic and technologically important Hamiltonian.
The solid electrolyte $\beta$-alumina is known for its high ionic conductivity~\cite{aluminas_conductivity, aluminas_structure} and there is broad general interest in using it for low-carbon energy storage~\cite{aluminas_battery}. 
Of particular interest for these applications is the accurate calculation of equilibrium voltage, ionic mobility, and thermal stability, which are all directly related to accurate ground state energies of the battery as outlined by Delgado \textit{et al.} in their work on the lithium-ion battery Li$_2$FeSiO$_4$~\cite{delgado2022Apr}.

Classical computation of accurate ground state energies of the $\beta$-aluminas is challenged by the non-stoichiometric chemical composition, Na$_{1+x}$Al$_{11}$O$_{17 + x/2}$, which requires large supercells to resolve finite-size effects.
This is a larger supercell than has been considered in other resource estimates of materials.
We choose this particular example because it is large enough that mean-field classical heuristics (e.g., density functional theory) are likely to be the only methods that are viable, potentially leading to small values of $\overlap_i$.
However, an analysis of the precise value of $\overlap_i$ that is classically achievable with these heuristics is beyond the scope of this Letter and thus we leave it as a free parameter.

\begin{figure}
\includegraphics{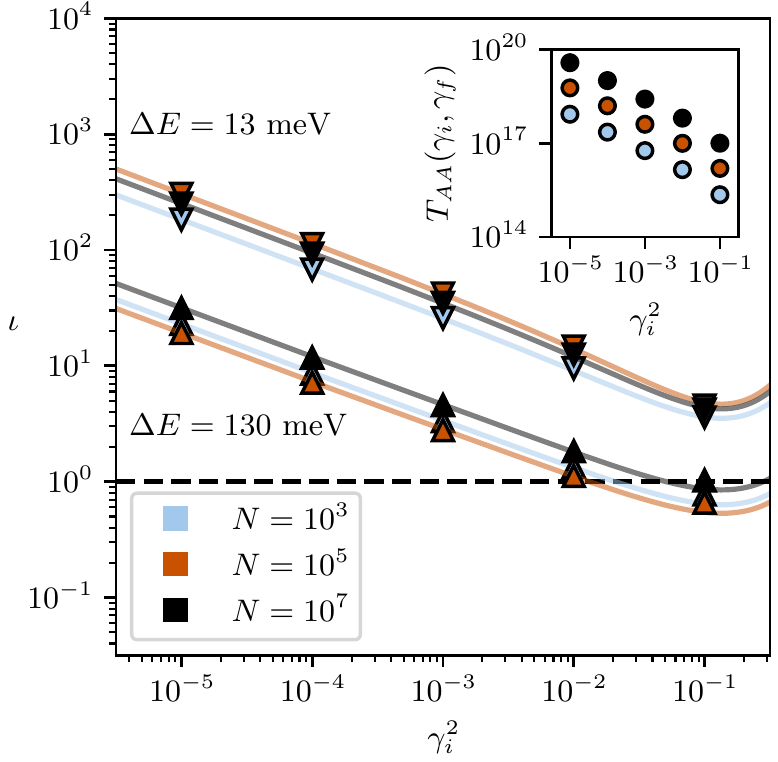}
\caption{$T$ counts and improvement $\iota$ for $\beta$-alumina structure Na$_{4}$Al$_{22}$O$_{35}$ with $\eta$ = 610, $N = 10^3, \ 10^5, \ 10^7$, with $\epsilon$ saturating the bound in Eq.~\ref{eq:approximation_bound} with $\overlap_f^2$ = 0.75, $\overlap_i$ ranging from $10^{-5}$ to $10^{-1}$ and two different values of $\Delta E$.
The inset shows $T_{AA}$ for various system sizes, with the main figure presenting $\iota$ with solid lines fit to the asymptotic form Eq.~\ref{eq:asymptotic_improvement}.
The dashed line is a guide for the eye at $\iota = 1$.
\label{fig:1quant_results}
}
\end{figure}

In Fig.~\ref{fig:1quant_results} we present resource estimates for a Na$_{4}$Al$_{22}$O$_{35}$ supercell with 610 electrons using a first quantized representation~\cite{babbush2019quantum,su2021fault} of the electronic structure Hamiltonian in a plane-wave basis set with cardinality $N$.
We find that $\iota > 1$ in all cases where $\overlap_i^2 \leq 10^{-2}$, from small to large basis sets, and for $\Delta E$ below chemical accuracy to different extents. 
We also see that the $T$ counts for $AA$ are such that it is plausible to imagine implementing calculations like this on fault-tolerant quantum computers that are perhaps a generation beyond the ones considered in Refs.~\cite{babbush2018encoding,lee2021even}.
Even when starting with a fairly large $\overlap_i^2 = 0.1$, we still find an order of magnitude improvement in T counts for a high accuracy calculation, as a result of investing resources in better state preparation.
We note that 33,275 logical qubits are required for the $N = 10^7$ calculation with $\Delta E = 13$ meV; 18,635 logical qubits for anti-symmetrization and 14,640 logical qubits for all other computations.
A detailed analysis of qubit counts is relegated to the SM~\cite{SMref}, as well as the full source code for computing resource estimates.

\textit{Conclusions.---}
We have developed resource estimates for end-to-end ground state energy determination using near-optimal state preparation~\cite{lin2020near}.
The ratio of the $T$ count for successful ground state energy estimation, without and with this state preparation, define an improvement factor that is related to likely runtime reductions.
This improvement is near-quadratic in $\overlap_i$ and demonstrated credible multiple-order-of-magnitude speedups for a toy problem and a highly realistic electronic structure problem.

Future work will involve determining more realistic estimates for scenarios under which these types of speedups will be realized.
In particular, categorizing the values of $\overlap_i$ typical of classical heuristics that are efficiently implementable as $\mathcal{U}_{\agsi}$ is an open research area.
It also remains unclear whether efficient implementations of adiabatic state preparation or other variants on filter-based state preparation are more or less efficient than the one examined in this Letter.
Finally, whether quantum phase estimation protocols with built-in tolerance to state preparation errors~\cite{russo2021evaluating,ding2022even} can be exploited to achieve better improvements is a topic for future work.

\textit{Note added.---} Between uploading the first and second versions of this Letter to the arXiv, we became aware of another manuscript considering similar aspects of ground state preparation~\cite{gratsea2022reject}.

\begin{acknowledgments}
We gratefully acknowledge useful conversations with
Ryan Babbush, 
Anand Ganti, 
Lucas Kocia, 
Alina Kononov, 
Michael Kreshchuk, 
Andrew Landahl, 
Lin Lin, 
Alicia Magann, 
Jonathan Moussa, 
Setso Metodi, 
Mason Rhodes, and 
Norm Tubman. 
All authors were supported by the National Nuclear Security Administration's Advanced Simulation and Computing Program.
AER was partially supported by the U.S. Department of Energy, Office of Science, Office of Advanced Scientific Computing Research, Quantum Computing Application Teams program.
ADB was partially supported by the U.S. Department of Energy, Office of Science, National Quantum Information Science Research Centers program and Sandia National Laboratories' Laboratory Directed Research and Development program (Project 222396).

This article has been co-authored by employees of National Technology \& Engineering Solutions of Sandia, LLC under Contract No. DE-NA0003525 with the U.S. Department of Energy (DOE). 
The authors own all right, title and interest in and to the article and are solely responsible for its contents. 
The United States Government retains and the publisher, by accepting the article for publication, acknowledges that the United States Government retains a non-exclusive, paid-up, irrevocable, world-wide license to publish or reproduce the published form of this article or allow others to do so, for United States Government purposes.
The DOE will provide public access to these results of federally sponsored research in accordance with the DOE Public Access Plan \url{https://www.energy.gov/downloads/doe-public-access-plan}.
\end{acknowledgments}

\bibliography{references}

\begin{thebibliography}{66}%
\makeatletter
\providecommand \@ifxundefined [1]{%
 \@ifx{#1\undefined}
}%
\providecommand \@ifnum [1]{%
 \ifnum #1\expandafter \@firstoftwo
 \else \expandafter \@secondoftwo
 \fi
}%
\providecommand \@ifx [1]{%
 \ifx #1\expandafter \@firstoftwo
 \else \expandafter \@secondoftwo
 \fi
}%
\providecommand \natexlab [1]{#1}%
\providecommand \enquote  [1]{``#1''}%
\providecommand \bibnamefont  [1]{#1}%
\providecommand \bibfnamefont [1]{#1}%
\providecommand \citenamefont [1]{#1}%
\providecommand \href@noop [0]{\@secondoftwo}%
\providecommand \href [0]{\begingroup \@sanitize@url \@href}%
\providecommand \@href[1]{\@@startlink{#1}\@@href}%
\providecommand \@@href[1]{\endgroup#1\@@endlink}%
\providecommand \@sanitize@url [0]{\catcode `\\12\catcode `\$12\catcode
  `\&12\catcode `\#12\catcode `\^12\catcode `\_12\catcode `\%12\relax}%
\providecommand \@@startlink[1]{}%
\providecommand \@@endlink[0]{}%
\providecommand \url  [0]{\begingroup\@sanitize@url \@url }%
\providecommand \@url [1]{\endgroup\@href {#1}{\urlprefix }}%
\providecommand \urlprefix  [0]{URL }%
\providecommand \Eprint [0]{\href }%
\providecommand \doibase [0]{http://dx.doi.org/}%
\providecommand \selectlanguage [0]{\@gobble}%
\providecommand \bibinfo  [0]{\@secondoftwo}%
\providecommand \bibfield  [0]{\@secondoftwo}%
\providecommand \translation [1]{[#1]}%
\providecommand \BibitemOpen [0]{}%
\providecommand \bibitemStop [0]{}%
\providecommand \bibitemNoStop [0]{.\EOS\space}%
\providecommand \EOS [0]{\spacefactor3000\relax}%
\providecommand \BibitemShut  [1]{\csname bibitem#1\endcsname}%
\let\auto@bib@innerbib\@empty
\bibitem [{\citenamefont {Lloyd}(1996)}]{lloyd1996universal}%
  \BibitemOpen
  \bibfield  {author} {\bibinfo {author} {\bibfnamefont {S.}~\bibnamefont
  {Lloyd}},\ }\href {\doibase 10.1126/science.273.5278.1073} {\bibfield
  {journal} {\bibinfo  {journal} {Science}\ }\textbf {\bibinfo {volume}
  {273}},\ \bibinfo {pages} {1073} (\bibinfo {year} {1996})}\BibitemShut
  {NoStop}%
\bibitem [{\citenamefont {Aspuru-Guzik}\ \emph {et~al.}(2005)\citenamefont
  {Aspuru-Guzik}, \citenamefont {Dutoi}, \citenamefont {Love},\ and\
  \citenamefont {Head-Gordon}}]{aspuru2005simulated}%
  \BibitemOpen
  \bibfield  {author} {\bibinfo {author} {\bibfnamefont {A.}~\bibnamefont
  {Aspuru-Guzik}}, \bibinfo {author} {\bibfnamefont {A.~D.}\ \bibnamefont
  {Dutoi}}, \bibinfo {author} {\bibfnamefont {P.~J.}\ \bibnamefont {Love}}, \
  and\ \bibinfo {author} {\bibfnamefont {M.}~\bibnamefont {Head-Gordon}},\
  }\href {\doibase 10.1126/science.1113479} {\bibfield  {journal} {\bibinfo
  {journal} {Science}\ }\textbf {\bibinfo {volume} {309}},\ \bibinfo {pages}
  {1704} (\bibinfo {year} {2005})}\BibitemShut {NoStop}%
\bibitem [{\citenamefont {Jordan}\ \emph {et~al.}(2012)\citenamefont {Jordan},
  \citenamefont {Lee},\ and\ \citenamefont {Preskill}}]{jordan2012quantum}%
  \BibitemOpen
  \bibfield  {author} {\bibinfo {author} {\bibfnamefont {S.~P.}\ \bibnamefont
  {Jordan}}, \bibinfo {author} {\bibfnamefont {K.~S.}\ \bibnamefont {Lee}}, \
  and\ \bibinfo {author} {\bibfnamefont {J.}~\bibnamefont {Preskill}},\ }\href
  {\doibase 10.1126/science.1217069} {\bibfield  {journal} {\bibinfo  {journal}
  {Science}\ }\textbf {\bibinfo {volume} {336}},\ \bibinfo {pages} {1130}
  (\bibinfo {year} {2012})}\BibitemShut {NoStop}%
\bibitem [{\citenamefont {Poulin}\ and\ \citenamefont
  {Wocjan}(2009)}]{poulin2009preparing}%
  \BibitemOpen
  \bibfield  {author} {\bibinfo {author} {\bibfnamefont {D.}~\bibnamefont
  {Poulin}}\ and\ \bibinfo {author} {\bibfnamefont {P.}~\bibnamefont
  {Wocjan}},\ }\href {\doibase 10.1103/PhysRevLett.102.130503} {\bibfield
  {journal} {\bibinfo  {journal} {Physical review letters}\ }\textbf {\bibinfo
  {volume} {102}},\ \bibinfo {pages} {130503} (\bibinfo {year}
  {2009})}\BibitemShut {NoStop}%
\bibitem [{\citenamefont {Tubman}\ \emph {et~al.}(2018)\citenamefont {Tubman},
  \citenamefont {Mejuto-Zaera}, \citenamefont {Epstein}, \citenamefont {Hait},
  \citenamefont {Levine}, \citenamefont {Huggins}, \citenamefont {Jiang},
  \citenamefont {McClean}, \citenamefont {Babbush}, \citenamefont {Head-Gordon}
  \emph {et~al.}}]{tubman2018postponing}%
  \BibitemOpen
  \bibfield  {author} {\bibinfo {author} {\bibfnamefont {N.~M.}\ \bibnamefont
  {Tubman}}, \bibinfo {author} {\bibfnamefont {C.}~\bibnamefont
  {Mejuto-Zaera}}, \bibinfo {author} {\bibfnamefont {J.~M.}\ \bibnamefont
  {Epstein}}, \bibinfo {author} {\bibfnamefont {D.}~\bibnamefont {Hait}},
  \bibinfo {author} {\bibfnamefont {D.~S.}\ \bibnamefont {Levine}}, \bibinfo
  {author} {\bibfnamefont {W.}~\bibnamefont {Huggins}}, \bibinfo {author}
  {\bibfnamefont {Z.}~\bibnamefont {Jiang}}, \bibinfo {author} {\bibfnamefont
  {J.~R.}\ \bibnamefont {McClean}}, \bibinfo {author} {\bibfnamefont
  {R.}~\bibnamefont {Babbush}}, \bibinfo {author} {\bibfnamefont
  {M.}~\bibnamefont {Head-Gordon}},  \emph {et~al.},\ }\href@noop {} {\bibfield
   {journal} {\bibinfo  {journal} {arXiv preprint arXiv:1809.05523}\ }
  (\bibinfo {year} {2018})}\BibitemShut {NoStop}%
\bibitem [{\citenamefont {Ge}\ \emph {et~al.}(2019)\citenamefont {Ge},
  \citenamefont {Tura},\ and\ \citenamefont {Cirac}}]{ge2019faster}%
  \BibitemOpen
  \bibfield  {author} {\bibinfo {author} {\bibfnamefont {Y.}~\bibnamefont
  {Ge}}, \bibinfo {author} {\bibfnamefont {J.}~\bibnamefont {Tura}}, \ and\
  \bibinfo {author} {\bibfnamefont {J.~I.}\ \bibnamefont {Cirac}},\ }\href
  {\doibase 10.1063/1.5027484} {\bibfield  {journal} {\bibinfo  {journal}
  {Journal of Mathematical Physics}\ }\textbf {\bibinfo {volume} {60}},\
  \bibinfo {pages} {022202} (\bibinfo {year} {2019})}\BibitemShut {NoStop}%
\bibitem [{\citenamefont {Lin}\ and\ \citenamefont
  {Tong}(2020{\natexlab{a}})}]{lin2020near}%
  \BibitemOpen
  \bibfield  {author} {\bibinfo {author} {\bibfnamefont {L.}~\bibnamefont
  {Lin}}\ and\ \bibinfo {author} {\bibfnamefont {Y.}~\bibnamefont {Tong}},\
  }\href {\doibase 10.22331/q-2020-12-14-372} {\bibfield  {journal} {\bibinfo
  {journal} {Quantum}\ }\textbf {\bibinfo {volume} {4}},\ \bibinfo {pages}
  {372} (\bibinfo {year} {2020}{\natexlab{a}})}\BibitemShut {NoStop}%
\bibitem [{\citenamefont {Lemieux}\ \emph {et~al.}(2021)\citenamefont
  {Lemieux}, \citenamefont {Duclos-Cianci}, \citenamefont {S{\'e}n{\'e}chal},\
  and\ \citenamefont {Poulin}}]{lemieux2021resource}%
  \BibitemOpen
  \bibfield  {author} {\bibinfo {author} {\bibfnamefont {J.}~\bibnamefont
  {Lemieux}}, \bibinfo {author} {\bibfnamefont {G.}~\bibnamefont
  {Duclos-Cianci}}, \bibinfo {author} {\bibfnamefont {D.}~\bibnamefont
  {S{\'e}n{\'e}chal}}, \ and\ \bibinfo {author} {\bibfnamefont
  {D.}~\bibnamefont {Poulin}},\ }\href {\doibase 10.1103/PhysRevA.103.052408}
  {\bibfield  {journal} {\bibinfo  {journal} {Physical Review A}\ }\textbf
  {\bibinfo {volume} {103}},\ \bibinfo {pages} {052408} (\bibinfo {year}
  {2021})}\BibitemShut {NoStop}%
\bibitem [{\citenamefont {Reiher}\ \emph {et~al.}(2017)\citenamefont {Reiher},
  \citenamefont {Wiebe}, \citenamefont {Svore}, \citenamefont {Wecker},\ and\
  \citenamefont {Troyer}}]{reiher2017elucidating}%
  \BibitemOpen
  \bibfield  {author} {\bibinfo {author} {\bibfnamefont {M.}~\bibnamefont
  {Reiher}}, \bibinfo {author} {\bibfnamefont {N.}~\bibnamefont {Wiebe}},
  \bibinfo {author} {\bibfnamefont {K.~M.}\ \bibnamefont {Svore}}, \bibinfo
  {author} {\bibfnamefont {D.}~\bibnamefont {Wecker}}, \ and\ \bibinfo {author}
  {\bibfnamefont {M.}~\bibnamefont {Troyer}},\ }\href {\doibase
  10.1073/pnas.1619152114} {\bibfield  {journal} {\bibinfo  {journal}
  {Proceedings of the national academy of sciences}\ }\textbf {\bibinfo
  {volume} {114}},\ \bibinfo {pages} {7555} (\bibinfo {year}
  {2017})}\BibitemShut {NoStop}%
\bibitem [{\citenamefont {Babbush}\ \emph {et~al.}(2018)\citenamefont
  {Babbush}, \citenamefont {Gidney}, \citenamefont {Berry}, \citenamefont
  {Wiebe}, \citenamefont {McClean}, \citenamefont {Paler}, \citenamefont
  {Fowler},\ and\ \citenamefont {Neven}}]{babbush2018encoding}%
  \BibitemOpen
  \bibfield  {author} {\bibinfo {author} {\bibfnamefont {R.}~\bibnamefont
  {Babbush}}, \bibinfo {author} {\bibfnamefont {C.}~\bibnamefont {Gidney}},
  \bibinfo {author} {\bibfnamefont {D.~W.}\ \bibnamefont {Berry}}, \bibinfo
  {author} {\bibfnamefont {N.}~\bibnamefont {Wiebe}}, \bibinfo {author}
  {\bibfnamefont {J.}~\bibnamefont {McClean}}, \bibinfo {author} {\bibfnamefont
  {A.}~\bibnamefont {Paler}}, \bibinfo {author} {\bibfnamefont
  {A.}~\bibnamefont {Fowler}}, \ and\ \bibinfo {author} {\bibfnamefont
  {H.}~\bibnamefont {Neven}},\ }\href {\doibase 10.1103/PhysRevX.8.041015}
  {\bibfield  {journal} {\bibinfo  {journal} {Physical Review X}\ }\textbf
  {\bibinfo {volume} {8}},\ \bibinfo {pages} {041015} (\bibinfo {year}
  {2018})}\BibitemShut {NoStop}%
\bibitem [{\citenamefont {Lee}\ \emph {et~al.}(2021)\citenamefont {Lee},
  \citenamefont {Berry}, \citenamefont {Gidney}, \citenamefont {Huggins},
  \citenamefont {McClean}, \citenamefont {Wiebe},\ and\ \citenamefont
  {Babbush}}]{lee2021even}%
  \BibitemOpen
  \bibfield  {author} {\bibinfo {author} {\bibfnamefont {J.}~\bibnamefont
  {Lee}}, \bibinfo {author} {\bibfnamefont {D.~W.}\ \bibnamefont {Berry}},
  \bibinfo {author} {\bibfnamefont {C.}~\bibnamefont {Gidney}}, \bibinfo
  {author} {\bibfnamefont {W.~J.}\ \bibnamefont {Huggins}}, \bibinfo {author}
  {\bibfnamefont {J.~R.}\ \bibnamefont {McClean}}, \bibinfo {author}
  {\bibfnamefont {N.}~\bibnamefont {Wiebe}}, \ and\ \bibinfo {author}
  {\bibfnamefont {R.}~\bibnamefont {Babbush}},\ }\href {\doibase
  10.1103/PRXQuantum.2.030305} {\bibfield  {journal} {\bibinfo  {journal} {PRX
  Quantum}\ }\textbf {\bibinfo {volume} {2}},\ \bibinfo {pages} {030305}
  (\bibinfo {year} {2021})}\BibitemShut {NoStop}%
\bibitem [{\citenamefont {Su}\ \emph {et~al.}(2021{\natexlab{a}})\citenamefont
  {Su}, \citenamefont {Berry}, \citenamefont {Wiebe}, \citenamefont {Rubin},\
  and\ \citenamefont {Babbush}}]{su2021fault}%
  \BibitemOpen
  \bibfield  {author} {\bibinfo {author} {\bibfnamefont {Y.}~\bibnamefont
  {Su}}, \bibinfo {author} {\bibfnamefont {D.~W.}\ \bibnamefont {Berry}},
  \bibinfo {author} {\bibfnamefont {N.}~\bibnamefont {Wiebe}}, \bibinfo
  {author} {\bibfnamefont {N.}~\bibnamefont {Rubin}}, \ and\ \bibinfo {author}
  {\bibfnamefont {R.}~\bibnamefont {Babbush}},\ }\href {\doibase
  10.1103/PRXQuantum.2.040332} {\bibfield  {journal} {\bibinfo  {journal} {PRX
  Quantum}\ }\textbf {\bibinfo {volume} {2}},\ \bibinfo {pages} {040332}
  (\bibinfo {year} {2021}{\natexlab{a}})}\BibitemShut {NoStop}%
\bibitem [{\citenamefont {Su}\ \emph {et~al.}(2021{\natexlab{b}})\citenamefont
  {Su}, \citenamefont {Huang},\ and\ \citenamefont {Campbell}}]{su2021nearly}%
  \BibitemOpen
  \bibfield  {author} {\bibinfo {author} {\bibfnamefont {Y.}~\bibnamefont
  {Su}}, \bibinfo {author} {\bibfnamefont {H.-Y.}\ \bibnamefont {Huang}}, \
  and\ \bibinfo {author} {\bibfnamefont {E.~T.}\ \bibnamefont {Campbell}},\
  }\href {\doibase 10.22331/q-2021-07-05-495} {\bibfield  {journal} {\bibinfo
  {journal} {Quantum}\ }\textbf {\bibinfo {volume} {5}},\ \bibinfo {pages}
  {495} (\bibinfo {year} {2021}{\natexlab{b}})}\BibitemShut {NoStop}%
\bibitem [{\citenamefont {von Burg}\ \emph {et~al.}(2021)\citenamefont {von
  Burg}, \citenamefont {Low}, \citenamefont {H{\"a}ner}, \citenamefont
  {Steiger}, \citenamefont {Reiher}, \citenamefont {Roetteler},\ and\
  \citenamefont {Troyer}}]{von2021quantum}%
  \BibitemOpen
  \bibfield  {author} {\bibinfo {author} {\bibfnamefont {V.}~\bibnamefont {von
  Burg}}, \bibinfo {author} {\bibfnamefont {G.~H.}\ \bibnamefont {Low}},
  \bibinfo {author} {\bibfnamefont {T.}~\bibnamefont {H{\"a}ner}}, \bibinfo
  {author} {\bibfnamefont {D.~S.}\ \bibnamefont {Steiger}}, \bibinfo {author}
  {\bibfnamefont {M.}~\bibnamefont {Reiher}}, \bibinfo {author} {\bibfnamefont
  {M.}~\bibnamefont {Roetteler}}, \ and\ \bibinfo {author} {\bibfnamefont
  {M.}~\bibnamefont {Troyer}},\ }\href {\doibase
  10.1103/PhysRevResearch.3.033055} {\bibfield  {journal} {\bibinfo  {journal}
  {Physical Review Research}\ }\textbf {\bibinfo {volume} {3}},\ \bibinfo
  {pages} {033055} (\bibinfo {year} {2021})}\BibitemShut {NoStop}%
\bibitem [{\citenamefont {Childs}\ \emph {et~al.}(2018)\citenamefont {Childs},
  \citenamefont {Maslov}, \citenamefont {Nam}, \citenamefont {Ross},\ and\
  \citenamefont {Su}}]{childs2018toward}%
  \BibitemOpen
  \bibfield  {author} {\bibinfo {author} {\bibfnamefont {A.~M.}\ \bibnamefont
  {Childs}}, \bibinfo {author} {\bibfnamefont {D.}~\bibnamefont {Maslov}},
  \bibinfo {author} {\bibfnamefont {Y.}~\bibnamefont {Nam}}, \bibinfo {author}
  {\bibfnamefont {N.~J.}\ \bibnamefont {Ross}}, \ and\ \bibinfo {author}
  {\bibfnamefont {Y.}~\bibnamefont {Su}},\ }\href {\doibase
  10.1073/pnas.1801723115} {\bibfield  {journal} {\bibinfo  {journal}
  {Proceedings of the National Academy of Sciences}\ }\textbf {\bibinfo
  {volume} {115}},\ \bibinfo {pages} {9456} (\bibinfo {year}
  {2018})}\BibitemShut {NoStop}%
\bibitem [{\citenamefont {Childs}\ and\ \citenamefont
  {Su}(2019)}]{childs2019nearly}%
  \BibitemOpen
  \bibfield  {author} {\bibinfo {author} {\bibfnamefont {A.~M.}\ \bibnamefont
  {Childs}}\ and\ \bibinfo {author} {\bibfnamefont {Y.}~\bibnamefont {Su}},\
  }\href {\doibase 10.1103/PhysRevLett.123.050503} {\bibfield  {journal}
  {\bibinfo  {journal} {Physical review letters}\ }\textbf {\bibinfo {volume}
  {123}},\ \bibinfo {pages} {050503} (\bibinfo {year} {2019})}\BibitemShut
  {NoStop}%
\bibitem [{\citenamefont {Tran}\ \emph {et~al.}(2021)\citenamefont {Tran},
  \citenamefont {Su}, \citenamefont {Carney},\ and\ \citenamefont
  {Taylor}}]{tran2021faster}%
  \BibitemOpen
  \bibfield  {author} {\bibinfo {author} {\bibfnamefont {M.~C.}\ \bibnamefont
  {Tran}}, \bibinfo {author} {\bibfnamefont {Y.}~\bibnamefont {Su}}, \bibinfo
  {author} {\bibfnamefont {D.}~\bibnamefont {Carney}}, \ and\ \bibinfo {author}
  {\bibfnamefont {J.~M.}\ \bibnamefont {Taylor}},\ }\href {\doibase
  10.1103/PRXQuantum.2.010323} {\bibfield  {journal} {\bibinfo  {journal} {PRX
  Quantum}\ }\textbf {\bibinfo {volume} {2}},\ \bibinfo {pages} {010323}
  (\bibinfo {year} {2021})}\BibitemShut {NoStop}%
\bibitem [{\citenamefont {Klco}\ and\ \citenamefont
  {Savage}(2019)}]{klco2019digitization}%
  \BibitemOpen
  \bibfield  {author} {\bibinfo {author} {\bibfnamefont {N.}~\bibnamefont
  {Klco}}\ and\ \bibinfo {author} {\bibfnamefont {M.~J.}\ \bibnamefont
  {Savage}},\ }\href {\doibase 10.1103/PhysRevA.99.052335} {\bibfield
  {journal} {\bibinfo  {journal} {Physical Review A}\ }\textbf {\bibinfo
  {volume} {99}},\ \bibinfo {pages} {052335} (\bibinfo {year}
  {2019})}\BibitemShut {NoStop}%
\bibitem [{\citenamefont {Lamm}\ \emph {et~al.}(2019)\citenamefont {Lamm},
  \citenamefont {Lawrence}, \citenamefont {Yamauchi}, \citenamefont
  {Collaboration} \emph {et~al.}}]{lamm2019general}%
  \BibitemOpen
  \bibfield  {author} {\bibinfo {author} {\bibfnamefont {H.}~\bibnamefont
  {Lamm}}, \bibinfo {author} {\bibfnamefont {S.}~\bibnamefont {Lawrence}},
  \bibinfo {author} {\bibfnamefont {Y.}~\bibnamefont {Yamauchi}}, \bibinfo
  {author} {\bibfnamefont {N.}~\bibnamefont {Collaboration}},  \emph {et~al.},\
  }\href {\doibase 10.1103/PhysRevD.100.034518} {\bibfield  {journal} {\bibinfo
   {journal} {Physical Review D}\ }\textbf {\bibinfo {volume} {100}},\ \bibinfo
  {pages} {034518} (\bibinfo {year} {2019})}\BibitemShut {NoStop}%
\bibitem [{\citenamefont {Shaw}\ \emph {et~al.}(2020)\citenamefont {Shaw},
  \citenamefont {Lougovski}, \citenamefont {Stryker},\ and\ \citenamefont
  {Wiebe}}]{shaw2020quantum}%
  \BibitemOpen
  \bibfield  {author} {\bibinfo {author} {\bibfnamefont {A.~F.}\ \bibnamefont
  {Shaw}}, \bibinfo {author} {\bibfnamefont {P.}~\bibnamefont {Lougovski}},
  \bibinfo {author} {\bibfnamefont {J.~R.}\ \bibnamefont {Stryker}}, \ and\
  \bibinfo {author} {\bibfnamefont {N.}~\bibnamefont {Wiebe}},\ }\href
  {\doibase 10.22331/q-2020-08-10-306} {\bibfield  {journal} {\bibinfo
  {journal} {Quantum}\ }\textbf {\bibinfo {volume} {4}},\ \bibinfo {pages}
  {306} (\bibinfo {year} {2020})}\BibitemShut {NoStop}%
\bibitem [{\citenamefont {Nielsen}\ and\ \citenamefont
  {Chuang}(2002)}]{nielsen2002quantum}%
  \BibitemOpen
  \bibfield  {author} {\bibinfo {author} {\bibfnamefont {M.~A.}\ \bibnamefont
  {Nielsen}}\ and\ \bibinfo {author} {\bibfnamefont {I.}~\bibnamefont
  {Chuang}},\ }\href@noop {} {\enquote {\bibinfo {title} {Quantum computation
  and quantum information},}\ } (\bibinfo {year} {2002})\BibitemShut {NoStop}%
\bibitem [{\citenamefont {Kitaev}\ \emph {et~al.}(2002)\citenamefont {Kitaev},
  \citenamefont {Shen},\ and\ \citenamefont {Vyalyi}}]{kitaev2002classical}%
  \BibitemOpen
  \bibfield  {author} {\bibinfo {author} {\bibfnamefont {A.~Y.}\ \bibnamefont
  {Kitaev}}, \bibinfo {author} {\bibfnamefont {A.}~\bibnamefont {Shen}}, \ and\
  \bibinfo {author} {\bibfnamefont {M.~N.}\ \bibnamefont {Vyalyi}},\ }\href
  {\doibase 10.1090/gsm/047} {\emph {\bibinfo {title} {Classical and quantum
  computation}}},\ \bibinfo {number} {47}\ (\bibinfo  {publisher} {American
  Mathematical Soc.},\ \bibinfo {year} {2002})\BibitemShut {NoStop}%
\bibitem [{\citenamefont {Kempe}\ \emph {et~al.}(2006)\citenamefont {Kempe},
  \citenamefont {Kitaev},\ and\ \citenamefont {Regev}}]{kempe2006complexity}%
  \BibitemOpen
  \bibfield  {author} {\bibinfo {author} {\bibfnamefont {J.}~\bibnamefont
  {Kempe}}, \bibinfo {author} {\bibfnamefont {A.}~\bibnamefont {Kitaev}}, \
  and\ \bibinfo {author} {\bibfnamefont {O.}~\bibnamefont {Regev}},\ }\href
  {\doibase 10.1137/S0097539704445226} {\bibfield  {journal} {\bibinfo
  {journal} {{SIAM} journal on computing}\ }\textbf {\bibinfo {volume} {35}},\
  \bibinfo {pages} {1070} (\bibinfo {year} {2006})}\BibitemShut {NoStop}%
\bibitem [{\citenamefont {Feynman}(1982)}]{feynman1982simulating}%
  \BibitemOpen
  \bibfield  {author} {\bibinfo {author} {\bibfnamefont {R.~P.}\ \bibnamefont
  {Feynman}},\ }\href@noop {} {\bibfield  {journal} {\bibinfo  {journal}
  {International Journal of Theoretical Physics}\ }\textbf {\bibinfo {volume}
  {21}},\ \bibinfo {pages} {467} (\bibinfo {year} {1982})}\BibitemShut
  {NoStop}%
\bibitem [{\citenamefont {O’Gorman}\ \emph {et~al.}(2022)\citenamefont
  {O’Gorman}, \citenamefont {Irani}, \citenamefont {Whitfield},\ and\
  \citenamefont {Fefferman}}]{ogorman2022intractability}%
  \BibitemOpen
  \bibfield  {author} {\bibinfo {author} {\bibfnamefont {B.}~\bibnamefont
  {O’Gorman}}, \bibinfo {author} {\bibfnamefont {S.}~\bibnamefont {Irani}},
  \bibinfo {author} {\bibfnamefont {J.}~\bibnamefont {Whitfield}}, \ and\
  \bibinfo {author} {\bibfnamefont {B.}~\bibnamefont {Fefferman}},\ }\href
  {\doibase 10.1103/PRXQuantum.3.020322} {\bibfield  {journal} {\bibinfo
  {journal} {PRX Quantum}\ }\textbf {\bibinfo {volume} {3}},\ \bibinfo {pages}
  {020322} (\bibinfo {year} {2022})}\BibitemShut {NoStop}%
\bibitem [{\citenamefont {Peruzzo}\ \emph {et~al.}(2014)\citenamefont
  {Peruzzo}, \citenamefont {McClean}, \citenamefont {Shadbolt}, \citenamefont
  {Yung}, \citenamefont {Zhou}, \citenamefont {Love}, \citenamefont
  {Aspuru-Guzik},\ and\ \citenamefont {O’brien}}]{peruzzo2014variational}%
  \BibitemOpen
  \bibfield  {author} {\bibinfo {author} {\bibfnamefont {A.}~\bibnamefont
  {Peruzzo}}, \bibinfo {author} {\bibfnamefont {J.}~\bibnamefont {McClean}},
  \bibinfo {author} {\bibfnamefont {P.}~\bibnamefont {Shadbolt}}, \bibinfo
  {author} {\bibfnamefont {M.-H.}\ \bibnamefont {Yung}}, \bibinfo {author}
  {\bibfnamefont {X.-Q.}\ \bibnamefont {Zhou}}, \bibinfo {author}
  {\bibfnamefont {P.~J.}\ \bibnamefont {Love}}, \bibinfo {author}
  {\bibfnamefont {A.}~\bibnamefont {Aspuru-Guzik}}, \ and\ \bibinfo {author}
  {\bibfnamefont {J.~L.}\ \bibnamefont {O’brien}},\ }\href {\doibase
  10.1038/ncomms5213} {\bibfield  {journal} {\bibinfo  {journal} {Nature
  communications}\ }\textbf {\bibinfo {volume} {5}},\ \bibinfo {pages} {1}
  (\bibinfo {year} {2014})}\BibitemShut {NoStop}%
\bibitem [{\citenamefont {O’Malley}\ \emph {et~al.}(2016)\citenamefont
  {O’Malley}, \citenamefont {Babbush}, \citenamefont {Kivlichan},
  \citenamefont {Romero}, \citenamefont {McClean}, \citenamefont {Barends},
  \citenamefont {Kelly}, \citenamefont {Roushan}, \citenamefont {Tranter},
  \citenamefont {Ding} \emph {et~al.}}]{omalley2016scalable}%
  \BibitemOpen
  \bibfield  {author} {\bibinfo {author} {\bibfnamefont {P.~J.}\ \bibnamefont
  {O’Malley}}, \bibinfo {author} {\bibfnamefont {R.}~\bibnamefont {Babbush}},
  \bibinfo {author} {\bibfnamefont {I.~D.}\ \bibnamefont {Kivlichan}}, \bibinfo
  {author} {\bibfnamefont {J.}~\bibnamefont {Romero}}, \bibinfo {author}
  {\bibfnamefont {J.~R.}\ \bibnamefont {McClean}}, \bibinfo {author}
  {\bibfnamefont {R.}~\bibnamefont {Barends}}, \bibinfo {author} {\bibfnamefont
  {J.}~\bibnamefont {Kelly}}, \bibinfo {author} {\bibfnamefont
  {P.}~\bibnamefont {Roushan}}, \bibinfo {author} {\bibfnamefont
  {A.}~\bibnamefont {Tranter}}, \bibinfo {author} {\bibfnamefont
  {N.}~\bibnamefont {Ding}},  \emph {et~al.},\ }\href {\doibase
  10.1103/PhysRevX.6.031007} {\bibfield  {journal} {\bibinfo  {journal}
  {Physical Review X}\ }\textbf {\bibinfo {volume} {6}},\ \bibinfo {pages}
  {031007} (\bibinfo {year} {2016})}\BibitemShut {NoStop}%
\bibitem [{Note1()}]{Note1}%
  \BibitemOpen
  \bibinfo {note} {This cost isn't always negligible. As has been noted in
  Ref.~\cite {delgado2022Apr}, sometimes even the cost of preparing simple
  input states can dominate simulation costs unless certain simplifying
  assumptions are made.}\BibitemShut {Stop}%
\bibitem [{\citenamefont {Kitaev}(1995)}]{kitaev1995quantum}%
  \BibitemOpen
  \bibfield  {author} {\bibinfo {author} {\bibfnamefont {A.~Y.}\ \bibnamefont
  {Kitaev}},\ }\href {\doibase 10.48550/arXiv.quant-ph/9511026} {\bibfield
  {journal} {\bibinfo  {journal} {arXiv preprint quant-ph/9511026}\ } (\bibinfo
  {year} {1995}),\ 10.48550/arXiv.quant-ph/9511026}\BibitemShut {NoStop}%
\bibitem [{\citenamefont {Kimmel}\ \emph {et~al.}(2015)\citenamefont {Kimmel},
  \citenamefont {Low},\ and\ \citenamefont {Yoder}}]{kimmel2015robust}%
  \BibitemOpen
  \bibfield  {author} {\bibinfo {author} {\bibfnamefont {S.}~\bibnamefont
  {Kimmel}}, \bibinfo {author} {\bibfnamefont {G.~H.}\ \bibnamefont {Low}}, \
  and\ \bibinfo {author} {\bibfnamefont {T.~J.}\ \bibnamefont {Yoder}},\ }\href
  {\doibase 10.1103/PhysRevA.92.062315} {\bibfield  {journal} {\bibinfo
  {journal} {Physical Review A}\ }\textbf {\bibinfo {volume} {92}},\ \bibinfo
  {pages} {062315} (\bibinfo {year} {2015})}\BibitemShut {NoStop}%
\bibitem [{\citenamefont {Wiebe}\ and\ \citenamefont
  {Granade}(2016)}]{wiebe2016efficient}%
  \BibitemOpen
  \bibfield  {author} {\bibinfo {author} {\bibfnamefont {N.}~\bibnamefont
  {Wiebe}}\ and\ \bibinfo {author} {\bibfnamefont {C.}~\bibnamefont
  {Granade}},\ }\href {\doibase 10.1103/PhysRevLett.117.010503} {\bibfield
  {journal} {\bibinfo  {journal} {Physical review letters}\ }\textbf {\bibinfo
  {volume} {117}},\ \bibinfo {pages} {010503} (\bibinfo {year}
  {2016})}\BibitemShut {NoStop}%
\bibitem [{\citenamefont {O’Brien}\ \emph {et~al.}(2021)\citenamefont
  {O’Brien}, \citenamefont {Polla}, \citenamefont {Rubin}, \citenamefont
  {Huggins}, \citenamefont {McArdle}, \citenamefont {Boixo}, \citenamefont
  {McClean},\ and\ \citenamefont {Babbush}}]{o2021error}%
  \BibitemOpen
  \bibfield  {author} {\bibinfo {author} {\bibfnamefont {T.~E.}\ \bibnamefont
  {O’Brien}}, \bibinfo {author} {\bibfnamefont {S.}~\bibnamefont {Polla}},
  \bibinfo {author} {\bibfnamefont {N.~C.}\ \bibnamefont {Rubin}}, \bibinfo
  {author} {\bibfnamefont {W.~J.}\ \bibnamefont {Huggins}}, \bibinfo {author}
  {\bibfnamefont {S.}~\bibnamefont {McArdle}}, \bibinfo {author} {\bibfnamefont
  {S.}~\bibnamefont {Boixo}}, \bibinfo {author} {\bibfnamefont {J.~R.}\
  \bibnamefont {McClean}}, \ and\ \bibinfo {author} {\bibfnamefont
  {R.}~\bibnamefont {Babbush}},\ }\href {\doibase 10.1103/PRXQuantum.2.020317}
  {\bibfield  {journal} {\bibinfo  {journal} {PRX Quantum}\ }\textbf {\bibinfo
  {volume} {2}},\ \bibinfo {pages} {020317} (\bibinfo {year}
  {2021})}\BibitemShut {NoStop}%
\bibitem [{\citenamefont {Russo}\ \emph {et~al.}(2021)\citenamefont {Russo},
  \citenamefont {Rudinger}, \citenamefont {Morrison},\ and\ \citenamefont
  {Baczewski}}]{russo2021evaluating}%
  \BibitemOpen
  \bibfield  {author} {\bibinfo {author} {\bibfnamefont {A.~E.}\ \bibnamefont
  {Russo}}, \bibinfo {author} {\bibfnamefont {K.~M.}\ \bibnamefont {Rudinger}},
  \bibinfo {author} {\bibfnamefont {B.~C.}\ \bibnamefont {Morrison}}, \ and\
  \bibinfo {author} {\bibfnamefont {A.~D.}\ \bibnamefont {Baczewski}},\ }\href
  {\doibase 10.1103/PhysRevLett.126.210501} {\bibfield  {journal} {\bibinfo
  {journal} {Physical Review Letters}\ }\textbf {\bibinfo {volume} {126}},\
  \bibinfo {pages} {210501} (\bibinfo {year} {2021})}\BibitemShut {NoStop}%
\bibitem [{\citenamefont {Lin}\ and\ \citenamefont
  {Tong}(2022)}]{lin2022heisenberg}%
  \BibitemOpen
  \bibfield  {author} {\bibinfo {author} {\bibfnamefont {L.}~\bibnamefont
  {Lin}}\ and\ \bibinfo {author} {\bibfnamefont {Y.}~\bibnamefont {Tong}},\
  }\href {\doibase 10.1103/PRXQuantum.3.010318} {\bibfield  {journal} {\bibinfo
   {journal} {PRX Quantum}\ }\textbf {\bibinfo {volume} {3}},\ \bibinfo {pages}
  {010318} (\bibinfo {year} {2022})}\BibitemShut {NoStop}%
\bibitem [{\citenamefont {Wang}\ \emph {et~al.}(2022)\citenamefont {Wang},
  \citenamefont {Stilck-Fran{\c{c}}a}, \citenamefont {Zhang}, \citenamefont
  {Zhu},\ and\ \citenamefont {Johnson}}]{wang2022quantum}%
  \BibitemOpen
  \bibfield  {author} {\bibinfo {author} {\bibfnamefont {G.}~\bibnamefont
  {Wang}}, \bibinfo {author} {\bibfnamefont {D.}~\bibnamefont
  {Stilck-Fran{\c{c}}a}}, \bibinfo {author} {\bibfnamefont {R.}~\bibnamefont
  {Zhang}}, \bibinfo {author} {\bibfnamefont {S.}~\bibnamefont {Zhu}}, \ and\
  \bibinfo {author} {\bibfnamefont {P.~D.}\ \bibnamefont {Johnson}},\ }\href
  {\doibase https://doi.org/10.48550/arXiv.2209.06811} {\bibfield  {journal}
  {\bibinfo  {journal} {arXiv preprint arXiv:2209.06811}\ } (\bibinfo {year}
  {2022}),\ https://doi.org/10.48550/arXiv.2209.06811}\BibitemShut {NoStop}%
\bibitem [{Note2()}]{Note2}%
  \BibitemOpen
  \bibinfo {note} {We expect that further improvements can be made by adopting
  the aforementioned ``more elaborate strategies'' but these are likely to be
  constant-factor improvements rather than asymptotic ones.}\BibitemShut
  {Stop}%
\bibitem [{\citenamefont {Farhi}\ \emph {et~al.}(2000)\citenamefont {Farhi},
  \citenamefont {Goldstone}, \citenamefont {Gutmann},\ and\ \citenamefont
  {Sipser}}]{farhi2000quantum}%
  \BibitemOpen
  \bibfield  {author} {\bibinfo {author} {\bibfnamefont {E.}~\bibnamefont
  {Farhi}}, \bibinfo {author} {\bibfnamefont {J.}~\bibnamefont {Goldstone}},
  \bibinfo {author} {\bibfnamefont {S.}~\bibnamefont {Gutmann}}, \ and\
  \bibinfo {author} {\bibfnamefont {M.}~\bibnamefont {Sipser}},\ }\href
  {\doibase 10.48550/arXiv.quant-ph/0001106} {\bibfield  {journal} {\bibinfo
  {journal} {arXiv preprint quant-ph/0001106}\ } (\bibinfo {year} {2000}),\
  10.48550/arXiv.quant-ph/0001106}\BibitemShut {NoStop}%
\bibitem [{\citenamefont {Albash}\ and\ \citenamefont
  {Lidar}(2018)}]{albash2018adiabatic}%
  \BibitemOpen
  \bibfield  {author} {\bibinfo {author} {\bibfnamefont {T.}~\bibnamefont
  {Albash}}\ and\ \bibinfo {author} {\bibfnamefont {D.~A.}\ \bibnamefont
  {Lidar}},\ }\href {\doibase 10.1103/RevModPhys.90.015002} {\bibfield
  {journal} {\bibinfo  {journal} {Reviews of Modern Physics}\ }\textbf
  {\bibinfo {volume} {90}},\ \bibinfo {pages} {015002} (\bibinfo {year}
  {2018})}\BibitemShut {NoStop}%
\bibitem [{\citenamefont {Wan}\ and\ \citenamefont {Kim}(2020)}]{wan2020fast}%
  \BibitemOpen
  \bibfield  {author} {\bibinfo {author} {\bibfnamefont {K.}~\bibnamefont
  {Wan}}\ and\ \bibinfo {author} {\bibfnamefont {I.}~\bibnamefont {Kim}},\
  }\href {\doibase 10.48550/arXiv.2004.04164} {\bibfield  {journal} {\bibinfo
  {journal} {arXiv preprint arXiv:2004.04164}\ } (\bibinfo {year} {2020}),\
  10.48550/arXiv.2004.04164}\BibitemShut {NoStop}%
\bibitem [{\citenamefont {Gily{\'e}n}\ \emph {et~al.}(2019)\citenamefont
  {Gily{\'e}n}, \citenamefont {Su}, \citenamefont {Low},\ and\ \citenamefont
  {Wiebe}}]{gilyen2019quantum}%
  \BibitemOpen
  \bibfield  {author} {\bibinfo {author} {\bibfnamefont {A.}~\bibnamefont
  {Gily{\'e}n}}, \bibinfo {author} {\bibfnamefont {Y.}~\bibnamefont {Su}},
  \bibinfo {author} {\bibfnamefont {G.~H.}\ \bibnamefont {Low}}, \ and\
  \bibinfo {author} {\bibfnamefont {N.}~\bibnamefont {Wiebe}},\ }in\ \href
  {\doibase 10.1145/3313276.3316366} {\emph {\bibinfo {booktitle} {Proceedings
  of the 51st Annual ACM SIGACT Symposium on Theory of Computing}}}\ (\bibinfo
  {year} {2019})\ pp.\ \bibinfo {pages} {193--204}\BibitemShut {NoStop}%
\bibitem [{\citenamefont {Lee}\ \emph {et~al.}(2022)\citenamefont {Lee},
  \citenamefont {Lee}, \citenamefont {Zhai}, \citenamefont {Tong},
  \citenamefont {Dalzell}, \citenamefont {Kumar}, \citenamefont {Helms},
  \citenamefont {Gray}, \citenamefont {Cui}, \citenamefont {Liu} \emph
  {et~al.}}]{lee2022is}%
  \BibitemOpen
  \bibfield  {author} {\bibinfo {author} {\bibfnamefont {S.}~\bibnamefont
  {Lee}}, \bibinfo {author} {\bibfnamefont {J.}~\bibnamefont {Lee}}, \bibinfo
  {author} {\bibfnamefont {H.}~\bibnamefont {Zhai}}, \bibinfo {author}
  {\bibfnamefont {Y.}~\bibnamefont {Tong}}, \bibinfo {author} {\bibfnamefont
  {A.~M.}\ \bibnamefont {Dalzell}}, \bibinfo {author} {\bibfnamefont
  {A.}~\bibnamefont {Kumar}}, \bibinfo {author} {\bibfnamefont
  {P.}~\bibnamefont {Helms}}, \bibinfo {author} {\bibfnamefont
  {J.}~\bibnamefont {Gray}}, \bibinfo {author} {\bibfnamefont {Z.-H.}\
  \bibnamefont {Cui}}, \bibinfo {author} {\bibfnamefont {W.}~\bibnamefont
  {Liu}},  \emph {et~al.},\ }\href@noop {} {\bibfield  {journal} {\bibinfo
  {journal} {arXiv preprint arXiv:2208.02199}\ } (\bibinfo {year}
  {2022})}\BibitemShut {NoStop}%
\bibitem [{\citenamefont {Fowler}\ \emph {et~al.}(2012)\citenamefont {Fowler},
  \citenamefont {Mariantoni}, \citenamefont {Martinis},\ and\ \citenamefont
  {Cleland}}]{fowler2012surface}%
  \BibitemOpen
  \bibfield  {author} {\bibinfo {author} {\bibfnamefont {A.~G.}\ \bibnamefont
  {Fowler}}, \bibinfo {author} {\bibfnamefont {M.}~\bibnamefont {Mariantoni}},
  \bibinfo {author} {\bibfnamefont {J.~M.}\ \bibnamefont {Martinis}}, \ and\
  \bibinfo {author} {\bibfnamefont {A.~N.}\ \bibnamefont {Cleland}},\ }\href
  {\doibase 10.1103/PhysRevA.86.032324} {\bibfield  {journal} {\bibinfo
  {journal} {Physical Review A}\ }\textbf {\bibinfo {volume} {86}},\ \bibinfo
  {pages} {032324} (\bibinfo {year} {2012})}\BibitemShut {NoStop}%
\bibitem [{\citenamefont {Litinski}(2019)}]{litinski2019magic}%
  \BibitemOpen
  \bibfield  {author} {\bibinfo {author} {\bibfnamefont {D.}~\bibnamefont
  {Litinski}},\ }\href {\doibase 10.22331/q-2019-12-02-205} {\bibfield
  {journal} {\bibinfo  {journal} {Quantum}\ }\textbf {\bibinfo {volume} {3}},\
  \bibinfo {pages} {205} (\bibinfo {year} {2019})}\BibitemShut {NoStop}%
\bibitem [{\citenamefont {Dong}\ \emph {et~al.}(2021)\citenamefont {Dong},
  \citenamefont {Meng}, \citenamefont {Whaley},\ and\ \citenamefont
  {Lin}}]{dong2021efficient}%
  \BibitemOpen
  \bibfield  {author} {\bibinfo {author} {\bibfnamefont {Y.}~\bibnamefont
  {Dong}}, \bibinfo {author} {\bibfnamefont {X.}~\bibnamefont {Meng}}, \bibinfo
  {author} {\bibfnamefont {K.~B.}\ \bibnamefont {Whaley}}, \ and\ \bibinfo
  {author} {\bibfnamefont {L.}~\bibnamefont {Lin}},\ }\href {\doibase
  10.1103/PhysRevA.103.042419} {\bibfield  {journal} {\bibinfo  {journal}
  {Physical Review A}\ }\textbf {\bibinfo {volume} {103}},\ \bibinfo {pages}
  {042419} (\bibinfo {year} {2021})}\BibitemShut {NoStop}%
\bibitem [{\citenamefont {Grover}(1996)}]{grover1996fast}%
  \BibitemOpen
  \bibfield  {author} {\bibinfo {author} {\bibfnamefont {L.~K.}\ \bibnamefont
  {Grover}},\ }in\ \href {\doibase 10.1145/237814.237866} {\emph {\bibinfo
  {booktitle} {Proceedings of the twenty-eighth annual ACM symposium on Theory
  of computing}}}\ (\bibinfo {year} {1996})\ pp.\ \bibinfo {pages}
  {212--219}\BibitemShut {NoStop}%
\bibitem [{\citenamefont {Brassard}\ \emph {et~al.}(2002)\citenamefont
  {Brassard}, \citenamefont {Hoyer}, \citenamefont {Mosca},\ and\ \citenamefont
  {Tapp}}]{brassard2002quantum}%
  \BibitemOpen
  \bibfield  {author} {\bibinfo {author} {\bibfnamefont {G.}~\bibnamefont
  {Brassard}}, \bibinfo {author} {\bibfnamefont {P.}~\bibnamefont {Hoyer}},
  \bibinfo {author} {\bibfnamefont {M.}~\bibnamefont {Mosca}}, \ and\ \bibinfo
  {author} {\bibfnamefont {A.}~\bibnamefont {Tapp}},\ }\href {\doibase
  10.1090/conm/305/05215} {\bibfield  {journal} {\bibinfo  {journal}
  {Contemporary Mathematics}\ }\textbf {\bibinfo {volume} {305}},\ \bibinfo
  {pages} {53} (\bibinfo {year} {2002})}\BibitemShut {NoStop}%
\bibitem [{SMr()}]{SMref}%
  \BibitemOpen
  \href@noop {} {}\bibinfo {note} {See Supplemental Materials for more
  details.}\BibitemShut {Stop}%
\bibitem [{\citenamefont {Low}\ and\ \citenamefont
  {Chuang}(2017)}]{low2017optimal}%
  \BibitemOpen
  \bibfield  {author} {\bibinfo {author} {\bibfnamefont {G.~H.}\ \bibnamefont
  {Low}}\ and\ \bibinfo {author} {\bibfnamefont {I.~L.}\ \bibnamefont
  {Chuang}},\ }\href {\doibase 10.1103/PhysRevLett.118.010501} {\bibfield
  {journal} {\bibinfo  {journal} {Physical review letters}\ }\textbf {\bibinfo
  {volume} {118}},\ \bibinfo {pages} {010501} (\bibinfo {year}
  {2017})}\BibitemShut {NoStop}%
\bibitem [{\citenamefont {Kocia}\ \emph {et~al.}(2022)\citenamefont {Kocia},
  \citenamefont {Calderon-Vargas}, \citenamefont {Grace}, \citenamefont
  {Magann}, \citenamefont {Larsen}, \citenamefont {Baczewski},\ and\
  \citenamefont {Sarovar}}]{kocia2022digital}%
  \BibitemOpen
  \bibfield  {author} {\bibinfo {author} {\bibfnamefont {L.}~\bibnamefont
  {Kocia}}, \bibinfo {author} {\bibfnamefont {F.~A.}\ \bibnamefont
  {Calderon-Vargas}}, \bibinfo {author} {\bibfnamefont {M.~D.}\ \bibnamefont
  {Grace}}, \bibinfo {author} {\bibfnamefont {A.~B.}\ \bibnamefont {Magann}},
  \bibinfo {author} {\bibfnamefont {J.~B.}\ \bibnamefont {Larsen}}, \bibinfo
  {author} {\bibfnamefont {A.~D.}\ \bibnamefont {Baczewski}}, \ and\ \bibinfo
  {author} {\bibfnamefont {M.}~\bibnamefont {Sarovar}},\ }\href {\doibase
  10.48550/arXiv.2209.06242} {\bibfield  {journal} {\bibinfo  {journal} {arXiv
  preprint arXiv:2209.06242}\ } (\bibinfo {year} {2022}),\
  10.48550/arXiv.2209.06242}\BibitemShut {NoStop}%
\bibitem [{\citenamefont {Stinchcombe}(1973)}]{tfimreview}%
  \BibitemOpen
  \bibfield  {author} {\bibinfo {author} {\bibfnamefont {R.~B.}\ \bibnamefont
  {Stinchcombe}},\ }\href {\doibase 10.1088/0022-3719/6/15/009} {\bibfield
  {journal} {\bibinfo  {journal} {Journal of Physics C: Solid State Physics}\
  }\textbf {\bibinfo {volume} {6}},\ \bibinfo {pages} {2459} (\bibinfo {year}
  {1973})}\BibitemShut {NoStop}%
\bibitem [{Note3()}]{Note3}%
  \BibitemOpen
  \bibinfo {note} {Open boundary conditions will have very similar
  costs.}\BibitemShut {Stop}%
\bibitem [{\citenamefont {Sivarajah}\ \emph {et~al.}(2020)\citenamefont
  {Sivarajah}, \citenamefont {Dilkes}, \citenamefont {Cowtan}, \citenamefont
  {Simmons}, \citenamefont {Edgington},\ and\ \citenamefont
  {Duncan}}]{sivarajah2020pytket}%
  \BibitemOpen
  \bibfield  {author} {\bibinfo {author} {\bibfnamefont {S.}~\bibnamefont
  {Sivarajah}}, \bibinfo {author} {\bibfnamefont {S.}~\bibnamefont {Dilkes}},
  \bibinfo {author} {\bibfnamefont {A.}~\bibnamefont {Cowtan}}, \bibinfo
  {author} {\bibfnamefont {W.}~\bibnamefont {Simmons}}, \bibinfo {author}
  {\bibfnamefont {A.}~\bibnamefont {Edgington}}, \ and\ \bibinfo {author}
  {\bibfnamefont {R.}~\bibnamefont {Duncan}},\ }\href {\doibase
  10.1088/2058-9565/ab8e92} {\bibfield  {journal} {\bibinfo  {journal} {Quantum
  Sci. Technol.}\ }\textbf {\bibinfo {volume} {6}},\ \bibinfo {pages} {014003}
  (\bibinfo {year} {2020})}\BibitemShut {NoStop}%
\bibitem [{\citenamefont {Yao}\ and\ \citenamefont
  {Kummer}(1967)}]{aluminas_conductivity}%
  \BibitemOpen
  \bibfield  {author} {\bibinfo {author} {\bibfnamefont {Y.-F.~Y.}\
  \bibnamefont {Yao}}\ and\ \bibinfo {author} {\bibfnamefont {J.~T.}\
  \bibnamefont {Kummer}},\ }\href {\doibase
  https://doi.org/10.1016/0022-1902(67)80301-4} {\bibfield  {journal} {\bibinfo
   {journal} {Journal of Inorganic and Nuclear Chemistry}\ }\textbf {\bibinfo
  {volume} {29}},\ \bibinfo {pages} {2453} (\bibinfo {year}
  {1967})}\BibitemShut {NoStop}%
\bibitem [{\citenamefont {Stevens}\ and\ \citenamefont
  {Binner}(1984)}]{aluminas_structure}%
  \BibitemOpen
  \bibfield  {author} {\bibinfo {author} {\bibfnamefont {R.}~\bibnamefont
  {Stevens}}\ and\ \bibinfo {author} {\bibfnamefont {J.~G.~P.}\ \bibnamefont
  {Binner}},\ }\href {\doibase 10.1007/BF00540440} {\bibfield  {journal}
  {\bibinfo  {journal} {Journal of Materials Science}\ }\textbf {\bibinfo
  {volume} {19}},\ \bibinfo {pages} {695} (\bibinfo {year} {1984})}\BibitemShut
  {NoStop}%
\bibitem [{\citenamefont {Collongues}\ \emph {et~al.}(1984)\citenamefont
  {Collongues}, \citenamefont {Gourier}, \citenamefont {Kahn}, \citenamefont
  {Boilot}, \citenamefont {Colomban},\ and\ \citenamefont
  {Wicker}}]{aluminas_battery}%
  \BibitemOpen
  \bibfield  {author} {\bibinfo {author} {\bibfnamefont {R.}~\bibnamefont
  {Collongues}}, \bibinfo {author} {\bibfnamefont {D.}~\bibnamefont {Gourier}},
  \bibinfo {author} {\bibfnamefont {A.}~\bibnamefont {Kahn}}, \bibinfo {author}
  {\bibfnamefont {J.}~\bibnamefont {Boilot}}, \bibinfo {author} {\bibfnamefont
  {P.}~\bibnamefont {Colomban}}, \ and\ \bibinfo {author} {\bibfnamefont
  {A.}~\bibnamefont {Wicker}},\ }\href {\doibase
  https://doi.org/10.1016/0022-3697(84)90045-3} {\bibfield  {journal} {\bibinfo
   {journal} {Journal of Physics and Chemistry of Solids}\ }\textbf {\bibinfo
  {volume} {45}},\ \bibinfo {pages} {981} (\bibinfo {year} {1984})}\BibitemShut
  {NoStop}%
\bibitem [{\citenamefont {Delgado}\ \emph {et~al.}(2022)\citenamefont
  {Delgado}, \citenamefont {Casares}, \citenamefont {Reis}, \citenamefont
  {Zini}, \citenamefont {Campos}, \citenamefont
  {Cruz-Hern{\ifmmode\acute{a}\else\'{a}\fi}ndez}, \citenamefont {Voigt},
  \citenamefont {Lowe}, \citenamefont {Jahangiri}, \citenamefont
  {Martin-Delgado}, \citenamefont {Mueller},\ and\ \citenamefont
  {Arrazola}}]{delgado2022Apr}%
  \BibitemOpen
  \bibfield  {author} {\bibinfo {author} {\bibfnamefont {A.}~\bibnamefont
  {Delgado}}, \bibinfo {author} {\bibfnamefont {P.~A.~M.}\ \bibnamefont
  {Casares}}, \bibinfo {author} {\bibfnamefont {R.~d.}\ \bibnamefont {Reis}},
  \bibinfo {author} {\bibfnamefont {M.~S.}\ \bibnamefont {Zini}}, \bibinfo
  {author} {\bibfnamefont {R.}~\bibnamefont {Campos}}, \bibinfo {author}
  {\bibfnamefont {N.}~\bibnamefont
  {Cruz-Hern{\ifmmode\acute{a}\else\'{a}\fi}ndez}}, \bibinfo {author}
  {\bibfnamefont {A.-C.}\ \bibnamefont {Voigt}}, \bibinfo {author}
  {\bibfnamefont {A.}~\bibnamefont {Lowe}}, \bibinfo {author} {\bibfnamefont
  {S.}~\bibnamefont {Jahangiri}}, \bibinfo {author} {\bibfnamefont {M.~A.}\
  \bibnamefont {Martin-Delgado}}, \bibinfo {author} {\bibfnamefont {J.~E.}\
  \bibnamefont {Mueller}}, \ and\ \bibinfo {author} {\bibfnamefont {J.~M.}\
  \bibnamefont {Arrazola}},\ }\href {\doibase 10.48550/arXiv.2204.11890}
  {\bibfield  {journal} {\bibinfo  {journal} {arXiv}\ } (\bibinfo {year}
  {2022}),\ 10.48550/arXiv.2204.11890},\ \Eprint
  {http://arxiv.org/abs/2204.11890} {2204.11890} \BibitemShut {NoStop}%
\bibitem [{\citenamefont {Babbush}\ \emph {et~al.}(2019)\citenamefont
  {Babbush}, \citenamefont {Berry}, \citenamefont {McClean},\ and\
  \citenamefont {Neven}}]{babbush2019quantum}%
  \BibitemOpen
  \bibfield  {author} {\bibinfo {author} {\bibfnamefont {R.}~\bibnamefont
  {Babbush}}, \bibinfo {author} {\bibfnamefont {D.~W.}\ \bibnamefont {Berry}},
  \bibinfo {author} {\bibfnamefont {J.~R.}\ \bibnamefont {McClean}}, \ and\
  \bibinfo {author} {\bibfnamefont {H.}~\bibnamefont {Neven}},\ }\href
  {\doibase 10.1038/s41534-019-0199-y} {\bibfield  {journal} {\bibinfo
  {journal} {npj Quantum Information}\ }\textbf {\bibinfo {volume} {5}},\
  \bibinfo {pages} {1} (\bibinfo {year} {2019})}\BibitemShut {NoStop}%
\bibitem [{\citenamefont {Ding}\ and\ \citenamefont
  {Lin}(2022)}]{ding2022even}%
  \BibitemOpen
  \bibfield  {author} {\bibinfo {author} {\bibfnamefont {Z.}~\bibnamefont
  {Ding}}\ and\ \bibinfo {author} {\bibfnamefont {L.}~\bibnamefont {Lin}},\
  }\href {\doibase 10.48550/arXiv.2211.11973} {\bibfield  {journal} {\bibinfo
  {journal} {arXiv preprint arXiv:2211.11973}\ } (\bibinfo {year} {2022}),\
  10.48550/arXiv.2211.11973}\BibitemShut {NoStop}%
\bibitem [{\citenamefont {Gratsea}\ \emph {et~al.}(2022)\citenamefont
  {Gratsea}, \citenamefont {Sun},\ and\ \citenamefont
  {Johnson}}]{gratsea2022reject}%
  \BibitemOpen
  \bibfield  {author} {\bibinfo {author} {\bibfnamefont {K.}~\bibnamefont
  {Gratsea}}, \bibinfo {author} {\bibfnamefont {C.}~\bibnamefont {Sun}}, \ and\
  \bibinfo {author} {\bibfnamefont {P.~D.}\ \bibnamefont {Johnson}},\ }\href
  {\doibase 10.48550/arXiv.2212.09492} {\bibfield  {journal} {\bibinfo
  {journal} {arXiv preprint arXiv:2212.09492}\ } (\bibinfo {year} {2022}),\
  10.48550/arXiv.2212.09492}\BibitemShut {NoStop}%
\bibitem [{\citenamefont {Lin}\ and\ \citenamefont
  {Tong}(2020{\natexlab{b}})}]{lin2020optimalpolynomial}%
  \BibitemOpen
  \bibfield  {author} {\bibinfo {author} {\bibfnamefont {L.}~\bibnamefont
  {Lin}}\ and\ \bibinfo {author} {\bibfnamefont {Y.}~\bibnamefont {Tong}},\
  }\href {\doibase 10.22331/q-2020-11-11-361} {\bibfield  {journal} {\bibinfo
  {journal} {Quantum}\ }\textbf {\bibinfo {volume} {4}},\ \bibinfo {pages}
  {361} (\bibinfo {year} {2020}{\natexlab{b}})},\ \Eprint
  {http://arxiv.org/abs/1910.14596v4} {1910.14596v4} \BibitemShut {NoStop}%
\bibitem [{\citenamefont {Childs}\ and\ \citenamefont
  {Wiebe}(2012)}]{childs2012lcu}%
  \BibitemOpen
  \bibfield  {author} {\bibinfo {author} {\bibfnamefont {A.~M.}\ \bibnamefont
  {Childs}}\ and\ \bibinfo {author} {\bibfnamefont {N.}~\bibnamefont {Wiebe}},\
  }\href {\doibase 10.26421/qic12.11-12} {\bibfield  {journal} {\bibinfo
  {journal} {Quantum Information and Computation}\ }\textbf {\bibinfo {volume}
  {12}},\ \bibinfo {pages} {0901} (\bibinfo {year} {2012})},\ \Eprint
  {http://arxiv.org/abs/1202.5822} {1202.5822} \BibitemShut {NoStop}%
\bibitem [{\citenamefont {Selinger}(2015)}]{selingerR}%
  \BibitemOpen
  \bibfield  {author} {\bibinfo {author} {\bibfnamefont {P.}~\bibnamefont
  {Selinger}},\ }\href@noop {} {\bibfield  {journal} {\bibinfo  {journal}
  {Quantum Info. Comput.}\ }\textbf {\bibinfo {volume} {15}},\ \bibinfo {pages}
  {159–180} (\bibinfo {year} {2015})}\BibitemShut {NoStop}%
\bibitem [{\citenamefont {Niemann}\ \emph {et~al.}(2020)\citenamefont
  {Niemann}, \citenamefont {Wille},\ and\ \citenamefont
  {Drechsler}}]{niemannMCX}%
  \BibitemOpen
  \bibfield  {author} {\bibinfo {author} {\bibfnamefont {P.}~\bibnamefont
  {Niemann}}, \bibinfo {author} {\bibfnamefont {R.}~\bibnamefont {Wille}}, \
  and\ \bibinfo {author} {\bibfnamefont {R.}~\bibnamefont {Drechsler}},\ }\href
  {\doibase 10.1007/s11128-020-02816-0} {\bibfield  {journal} {\bibinfo
  {journal} {Quantum Information Processing}\ }\textbf {\bibinfo {volume}
  {19}},\ \bibinfo {pages} {317} (\bibinfo {year} {2020})}\BibitemShut
  {NoStop}%
\bibitem [{\citenamefont {Berry}\ \emph {et~al.}(2018)\citenamefont {Berry},
  \citenamefont {Kieferov{\'a}}, \citenamefont {Scherer}, \citenamefont
  {Sanders}, \citenamefont {Low}, \citenamefont {Wiebe}, \citenamefont
  {Gidney},\ and\ \citenamefont {Babbush}}]{berry2018improved}%
  \BibitemOpen
  \bibfield  {author} {\bibinfo {author} {\bibfnamefont {D.~W.}\ \bibnamefont
  {Berry}}, \bibinfo {author} {\bibfnamefont {M.}~\bibnamefont
  {Kieferov{\'a}}}, \bibinfo {author} {\bibfnamefont {A.}~\bibnamefont
  {Scherer}}, \bibinfo {author} {\bibfnamefont {Y.~R.}\ \bibnamefont
  {Sanders}}, \bibinfo {author} {\bibfnamefont {G.~H.}\ \bibnamefont {Low}},
  \bibinfo {author} {\bibfnamefont {N.}~\bibnamefont {Wiebe}}, \bibinfo
  {author} {\bibfnamefont {C.}~\bibnamefont {Gidney}}, \ and\ \bibinfo {author}
  {\bibfnamefont {R.}~\bibnamefont {Babbush}},\ }\href {\doibase
  10.1038/s41534-018-0071-5} {\bibfield  {journal} {\bibinfo  {journal} {npj
  Quantum Information}\ }\textbf {\bibinfo {volume} {4}},\ \bibinfo {pages} {1}
  (\bibinfo {year} {2018})}\BibitemShut {NoStop}%
\bibitem [{Note4()}]{Note4}%
  \BibitemOpen
  \bibinfo {note} {Actually, the filtering step succeeds with probability
  $>1/2$, so on average the step needs to be performed twice.}\BibitemShut
  {Stop}%
\bibitem [{\citenamefont {Casares}\ \emph {et~al.}(2022)\citenamefont
  {Casares}, \citenamefont {Campos},\ and\ \citenamefont
  {Martin-Delgado}}]{casares2022tfermion}%
  \BibitemOpen
  \bibfield  {author} {\bibinfo {author} {\bibfnamefont {P.~A.~M.}\
  \bibnamefont {Casares}}, \bibinfo {author} {\bibfnamefont {R.}~\bibnamefont
  {Campos}}, \ and\ \bibinfo {author} {\bibfnamefont {M.~A.}\ \bibnamefont
  {Martin-Delgado}},\ }\href {\doibase 10.22331/q-2022-07-20-768} {\bibfield
  {journal} {\bibinfo  {journal} {{Quantum}}\ }\textbf {\bibinfo {volume}
  {6}},\ \bibinfo {pages} {768} (\bibinfo {year} {2022})}\BibitemShut {NoStop}%
\end{thebibliography}%

\clearpage
\widetext
\begin{center}
\textbf{\large Supplemental Materials: \papertitle}
\end{center}

\setcounter{section}{0}
\setcounter{page}{1}

The Supplemental Materials elaborate on details of some of the central results in the main body of the paper.
\begin{itemize}
    \item Appendix A states and proves Theorem 1, which quantifies the impact of using an $\epsilon$ approximation to a reflector rather than an exact reflector in implementing amplitude amplification.
    \item Appendix B analyzes the impact of rotation synthesis errors.
    \item Appendix C describes the computation of $\mu$ and $\Delta$, including some details of the overhead associated with the implementation of Lin and Tong's binary search algorithm.
    \item Appendix D explores different choices for defining $\improvement$.
    \item Appendix E summarizes components of prior results from the literature that are essential to the synthesis in this paper.
\end{itemize}

\section{Appendix A: Theoretical analysis for bounds on \texorpdfstring{$\epsilon$}{ε}}
\setcounter{equation}{0}
\setcounter{figure}{0}
\renewcommand{\theequation}{A\arabic{equation}}
\renewcommand{\thefigure}{A\arabic{figure}}
\label{app:bounds}

Here we extend the error analysis of the filter-based ground state preparation technique developed by Lin and Tong \cite{lin2020near}.
In their work, they suggest state preparation using amplitude amplification, wherein a guess state $\agsi$ with overlap $\overlap_i$ to the true ground state $\gs$ of a system is boosted to a final state $\agsf$ with overlap $\overlap_f > \overlap_i$.
The problematic reflector required for this operation, a reflector around the true ground state $\mathcal{R}_{\gs}$, is implemented approximately in their work using quantum signal processing via an $\epsilon$-accurate block-encoded operator $\mathcal{R}_{\gs}(\epsilon)$.
In order to develop resources estimates for implementing Lin and Tong's algorithm, it is necessary to find a relationship between the approximation parameter $\epsilon$ and the quantities that define the amplitude amplification, namely $\overlap_i, \overlap_f$.
In this Appendix we compute a bound for $\epsilon$ such that state preparation using amplitude amplification with the approximate reflector $\mathcal{R}_{\gs}(\epsilon)$ is guaranteed to achieve a final overlap $\overlap_f$ starting from a guess state with overlap $\overlap_i$.

\subsection{Problem statement and theorem}
Given an initial state $\agsi$ with overlap $\overlap_i$ with the exact ground state $\gs$ of the Hamiltonian $\hamiltonian$, we would like to construct a state $\agsf$ which has an overlap $\overlap_f > \overlap_i$.
Our approach is to use quantum amplitude amplification, initially posited by Lin and Tong \cite{lin2020near}, wherein we boost the overlap by applying a sequence of two reflections:
\begin{equation}
    |\agsf \rangle = (\mathcal{R}_{\agsi}\mathcal{R}_{\gs})^{N_{iter}} |\agsi \rangle.
    \label{eq:qaa}
\end{equation}
Here $\mathcal{R}_{\agsi}$ is a reflector around $\agsi$, $\mathcal{R}_{\gs}$ a reflector around $\gs$ using quantum signal processing, and $N_{iter}$ the number of iterations which leads to the highest overlap defined in Eq~\ref{eq:niter}.

In practice one must approximate $\mathcal{R}_{\gs}$ as the exact ground state $\gs$ is the target one does not know, and wishes to construct.
As demonstrated by Lin and Tong~\cite{lin2020optimalpolynomial,lin2020near}, $\mathcal{R}_{\gs}$ can be approximately computed through a polynomial transformation of the Hamiltonian, constructed via quantum signal processing (QSP).
In their work, they first prove that there exists an approximation to the reflector as a polynomial function of $\hamiltonian$: $S(\frac{\hamiltonian - \mu I}{\alpha + |\mu|}, \epsilon, \delta)$, which is $\epsilon$-close in operator norm to $\mathcal{R}_{\gs}$ under the conditions $E_0 < \mu < E_1$ and $\delta \leq \min_k |E_k - \mu|/4\alpha$ where $E_k$ are the eigenvalues of $\hamiltonian$ and $\alpha \geq ||\hamiltonian||$.

The matrix function $S$, however, is not unitary and cannot be directly applied as a quantum circuit.
Instead, Lin and Tong describe a method to block-encode $S$ so that the fully block-ended unitary can then be implemented on a quantum computer.
They show that given an ($\alpha + |\mu|$, $m$, 0) block encoding of $\hamiltonian - \mu I$, one can construct a (1, $m+1$, 0) block-encoding of $S$ and thereby a (1, $m+1$, $\epsilon$) block-encoding of $\mathcal{R}_{\gs}$.
We will refer to the (1, $m+1$, $\epsilon$) block encoding of $\mathcal{R}_{\gs}$ hereafter as $\mathcal{R}_{\gs}(\epsilon)$, noting that the latter acts on both the state register storing $\agsi$ as we all as the auxiliary $m+1$ qubits required for block encoding..

The only leftover piece to complete the state preparation analysis suggested by Lin and Tong is determining what values of $\epsilon$ should be chosen to ensure that using $\mathcal{R}_{\gs}(\epsilon)$ within AA will yield a final overlap $\overlap_f$ starting with overlap $\overlap_i$.
In the next few sections we will show that a bound on $\epsilon$ can be analytically determined that satisfies the above conditions.
Below is a summary of our results:

\newtheorem{theorem}{Theorem}
\begin{theorem}
Consider a Hamiltonian $\hamiltonian$ with exact ground state $\gs$ and an initial guess state $\agsi$ with overlap $\overlap_i$.
The quantum amplitude amplification procedure $|\agsf\rangle = (\mathcal{R}_{\agsi} \mathcal{R}_{\gs}(\epsilon))^{N_{iter}}|\agsi\rangle$ described by initially by Lin and Tong~\cite{lin2020optimalpolynomial,lin2020near}
is guaranteed to return a state with, at minimum, a desired final overlap $\overlap_f$ if the following conditions are all met:

\begin{enumerate}
    \item $E_0 < \mu < E_1$, where $E_k$ are eigenvalues of $\hamiltonian$
    \item $\delta \leq \min_k|E_k - \mu|/4\alpha$
    \item $\epsilon \leq (1 - \overlap_f^2)/6N_{iter}^2$
\end{enumerate}
\end{theorem}

The first two conditions were already introduced in the work by Lin and Tong in their study of ground state preparation with quantum signal processing \cite{lin2020near}.
The last condition on $\epsilon$ has been introduced in this work, and what we will work towards proving in the next few sections.

It should be noted the form of amplitude amplification in Eq~\ref{eq:qaa} is different from the Brassard approach \cite{brassard2002quantum}.
We find, rather, that the approach in Eq~\ref{eq:qaa} yields a lower cost by nearly a factor of 2, as the QSP unitary only needs to be queried once per AA iteration, while in the Brassard AA approach the QSP unitary would need to be queried twice per iteration.
As the QSP query is the bulk of the cost in AA, this results in a significant reduction in circuit depth.

\subsection{Structure of the approximate reflector \texorpdfstring{$\mathcal{R}_{\gs}(\epsilon)$}{RΨ₀(ε)}}
In order to bound $\epsilon$ we will first study the structure and bounds of sub-matrices of $\mathcal{R}_{\gs}(\epsilon)$. 
To do so, we will first begin by assuming we are given a Hermitian block encoding of the Hamiltonian, which greatly simplifies all of the following analysis.
While this may not seem like a general assumption, Appendix C in Dong \textit{et al.} \cite{dong2021efficient} provides a technique for constructing a Hermitian block encoding from a non-Hermitian encoding with just a single-qubit overhead.

We begin with a few simple observations on the structure of $\mathcal{R}_{\gs}(\epsilon)$.

\begin{enumerate} 
\item $\mathcal{R}_{\gs}(\epsilon)$ operates identically and independently on states with support on the QSP $|0\rangle$ subspace and QSP $|1\rangle$ subspace.
We will work in the $|0\rangle$ subspace from here on out.

\item The matrix $\mathcal{R}_{\gs}(\epsilon)$ only contains real values as our polynomial function $S(\frac{\hamiltonian - \mu I}{\alpha + |\mu|}, \epsilon, \delta)$ is real. 
The mapping from the complex function that QSP implements to a real one is accomplished by adding a single ancilla qubit as described in Dong \textit{et al.} \cite{dong2021efficient}, Figure 3.

\item $\mathcal{R}_{\gs}(\epsilon)$ can be made symmetric.
To do so, we select symmetric phases as describe in Dong \textit{et al.} \cite{dong2021efficient}. Alongside the Hermiticity of $\mathcal{U}_\hamiltonian$ and realness of $\mathcal{R}_{\gs}(\epsilon)$, the symmetric phase factors guarantee that $\mathcal{R}_{\gs}(\epsilon)$ is a symmetric real matrix.

\end{enumerate}

With these three observations, we can write down the most general form of $\mathcal{R}_{\gs}(\epsilon)$ within the $|0\rangle$ QSP subspace as
\begin{equation}
\begin{split}
    &|0^m\rangle \ \  |\overline{0^m}\rangle \\
    \mathcal{R}_{\gs}(\epsilon) = &\begin{bmatrix}
    S(\epsilon) \ \ & A\ \ \\
    A^T \ \ & B\ \ \\
    \end{bmatrix}
\end{split}
\label{eq:fblock}
\end{equation}
where $S(\epsilon$) is shorthand for the Hamiltonian polynomial $S(\frac{\hamiltonian - \mu I}{\alpha + |\mu|}, \epsilon, \delta)$, $|0^{m}\rangle$ refers to the zeroed auxiliary register, $|\overline{0^m}\rangle$ its orthogonal complement, and two matrices $A, B$ with the latter being symmetric.
The matrices $A, B$ have no obvious internal structure, but we can prove bounds on the matrix norms of $A, B$ which will be integral in our analysis of the error propagation from $\epsilon$.

\subsection{Bounds on matrix norms of \texorpdfstring{$B, Q$}{B, Q}}
Our analysis of matrix norms relies on the unitarity of $\mathcal{R}_{\gs}(\epsilon)$.
Following the condition that $\mathcal{R}_{\gs}(\epsilon)^\dagger \mathcal{R}_{\gs}(\epsilon) = I$, we can derive three linearly independent conditions on matrix norms:
\begin{equation}
    S^2 + A^2 = I,\ SA^T + AB = 0,\ (A^T)^2 + B^2 = I
    \label{eq:matrixnorms}
\end{equation}

We know that $S$ is $\epsilon$ close in operator norm to the exact reflector $\mathcal{R}_{\gs}$, and more precisely has eigenvalues $\epsilon$ close to $\pm 1$\cite{lin2020near}.
As such, in the eigenbasis $S^2 = diag(\{(1 - \epsilon_i)^2\})$ where $\epsilon_i < \epsilon \ \forall \ i$ are the errors in each eigenvalue.
Considering the first equation in Eq~\ref{eq:matrixnorms}, we see that $A^2 = diag(\{(2\epsilon_i - \epsilon_i^2)\})$.
Since $\epsilon_i \leq \epsilon$, we can therefore bound the left and right as $||A^2|| \leq 2\epsilon - \epsilon^2$, which after taking the square root of both sides results in 
\begin{equation}
    ||A|| \leq \sqrt{2\epsilon - \epsilon^2}.
\label{eq:Qnorm}
\end{equation}
Bounds on $||B||$ are not required for our current analysis.
The only required information for this submatrix is that the norm is at most unity, in order to maintain unitarity.

\subsection{Bounds on overlap \texorpdfstring{$\overlap_f$}{γf}}
We can now work out bounds for the overlap after $N_{iter}$ applications of amplitude amplification. 
Again using the 2-dimensional basis as seen in Eq~\ref{eq:fblock}, we can write down the initial state as $\begin{pmatrix} \agsi & 0 \end{pmatrix}^T$, and the initial state reflector as 
\begin{equation}
    \begin{bmatrix}
    \mathcal{R}_{\agsi} & 0 \\
    0 & -I \\
    \end{bmatrix}
\end{equation}
We can then write out $(\mathcal{R}_{\agsi}\mathcal{R}_{\gs}(\epsilon))^k|\agsi\rangle$ for a few values of $k$, in this case for $k = 0, 1, 2, 3$:
\begin{equation}
\begin{split}
    \begin{bmatrix}
    \agsi \\ 0 \\
    \end{bmatrix},
    \begin{bmatrix}
    \mathcal{R}_{\agsi}S(\epsilon)\agsi \\ -A^T \agsi \\
    \end{bmatrix},
    \begin{bmatrix}
    (\mathcal{R}_{\agsi}S(\epsilon))^2\agsi - \mathcal{R}_{\agsi}(AA^T)\agsi \\ 
    A^T\mathcal{R}_{\agsi}S(\epsilon)\agsi + BA^T\agsi \\
    \end{bmatrix}\\
    \begin{bmatrix}
    (\mathcal{R}_{\agsi}S(\epsilon))^3\agsi -
    \mathcal{R}_{\agsi}S(\epsilon)\mathcal{R}_{\agsi}AA^T\agsi - 
    \mathcal{R}_{\agsi}AA^T\mathcal{R}_{\agsi}S(\epsilon)\agsi + 
    \mathcal{R}_{\agsi}ABA^T\agsi \\ 
    -A^T(\mathcal{R}_{\agsi}S(\epsilon))^2\agsi 
    + A^T \mathcal{R}_{\agsi}AA^T\agsi 
    + BA^T\mathcal{R}_{\agsi}S(\epsilon)\agsi - 
    B^2A^T\agsi\\ 
    \end{bmatrix}
\end{split}
\label{eq:exact_sequence}
\end{equation}
To reduce the above expressions to something reasonable, we can see that the vectors as a series can be written as: 
\begin{equation}
\Big[(\mathcal{R}_{\agsi}\mathcal{R}_{\gs}(\epsilon))^k|\agsi\rangle \Big]^T= 
\begin{bmatrix}
(\mathcal{R}_{\agsi}S(\epsilon))^k\agsi + a_k(\epsilon) &
b_k(\epsilon)
\end{bmatrix},
\label{eq:hilevel}
\end{equation}
where $a_k, b_k$ are some vectors that are functions of $\epsilon$ as presented in the explicitly computed sequence, and which depend on the iteration $k$.

We can now compute the overlap with the exact ground state using the expression in Eq~\ref{eq:hilevel}, and then reduce the solution using the bounds we determined in the previous section.
First, we look at the exact overlap
\begin{equation}
    \overlap_f^2 = \frac{|\langle \gs | (\mathcal{R}_{\agsi}S(\epsilon))^{N_{iter}} |\agsi \rangle + \langle \gs |a_{N_{iter}}(\epsilon) \rangle|^2}{|(\mathcal{R}_{\agsi}S(\epsilon))^{N_{iter}}\agsi+a_{N_{iter}}(\epsilon)|^2 + |b_{N_{iter}}(\epsilon)|^2}.
\end{equation}
At this stage the denominator is redundant, as the norm of the vector is always 1 due to unitarity, however we will be replacing $a_k, b_k$ soon to expression in lowest order to $\epsilon$, which will require proper normalization.

Now we can construct an approximate expression for $\overlap_f$ by computing each $\epsilon$ dependent term to lowest order in $\epsilon$.
The first quantity is $S(\epsilon)$, which is $\epsilon$-close in eigenvalues to the true reflection operator $\mathcal{R}_{\gs}$.
The next three quantities are the bounds on $a_k$, $b_k$ and $c_k$ which are determined by induction by looking at the series in Eq~\ref{eq:exact_sequence}: $|a_k(\epsilon)| \leq (k^2 - k)\epsilon + O(\epsilon^2)$, $|b_k(\epsilon)|\leq  k\sqrt{2\epsilon} + O(\epsilon)$.

Before stating our final result, we note that we want to ensure, under any condition, that $\overlap_f$ is achievable.
As such, we will write down the \textit{worst case} expression for $\overlap_f$ which is possible at the lowest order in $\epsilon$, allowing us to reduce certain factors like $|\langle\gs|(\mathcal{R}_{\agsi}S(\epsilon))^k|\agsi\rangle$ to 1 when multiplied by an error source.
With that being said, a final expression for $\overlap_f^2$, for the worst case in lowest order of $\epsilon$, is:
\begin{equation}
    \overlap_f^2 \geq |\langle \Psi_0 | (\mathcal{R}_{\agsi}R_{\Psi_0})^{N_{iter}} |\agsi\rangle|^2(1 - 6N_{iter}^2 \epsilon) + O(\epsilon^{3/2})
\label{eq:approximate_sf}
\end{equation}
Rearranging terms and noting that $|\langle \Psi_0 | (\mathcal{R}_{\agsi}R_{\Psi_0})^{N_{iter}} |\agsi\rangle|^2 = 1$, we then see that to ensure $\overlap_f$ final overlap we can pick 
\begin{equation}
    \epsilon \leq (1 - \overlap_f^2)/6N_{iter}^2.\ \ \square
\label{eq:final_epsilon}
\end{equation}
As a final remark, the $O(\epsilon^{3/2})$ term in Eq~\ref{eq:approximate_sf} actually increases the bound on $\epsilon$, as the next order corrections all emerge as $\sqrt{2\epsilon - \epsilon^2}$, effectively reducing the error from $b_k$.

\section{Appendix B: Impact of errors in rotation synthesis}
\setcounter{equation}{0}
\setcounter{figure}{0}
\renewcommand{\theequation}{B\arabic{equation}}
\renewcommand{\thefigure}{B\arabic{figure}}
\label{app:rotation_synthesis}

In the context of a fault-tolerant implementation of any quantum algorithm, the impact of approximation errors attendant to the synthesis of arbitrary-angle rotations from a finite gate set needs to be quantified.
There are two places where these rotation synthesis errors will occur in the algorithm under consideration in this Letter-- (1) the implementation of the Hamiltonian and (2) the implementation of the controlled QSP phase rotations $e^{i\phi_j \Pi}$.\aer{Rotation synthesis errors also occur in the Hartree-Fock initialization step.}
Source (1) is an aggregate truncation error associated with potentially numerous choices that are Hamiltonian- and implementation-dependent, which we quantify using a single parameter $\epsilon_{R,H}$.
Source (2) is a more straightforward truncation error that strictly depends on the precision with respect to which the QSP rotation angles are implemented, which we quantify using a second parameter $\epsilon_{R,P}$.
Here we describe the impact that these sources of error have on the accuracy and cost of state preparation using QSP at leading order in $\epsilon_{R,H}$ and $\epsilon_{R,P}$.
In the following analysis we assume that both of these sources of error are small relative to the QSP errors analyzed in Appendix A.

\subsection{Rotation synthesis error in \texorpdfstring{$\hamiltonian$}{H}}
For this analysis we consider an LCU implementation of the Hamiltonian, $\hamiltonian = \sum_{l = 0}^L w_l \hamiltonian_l / \alpha$, $\alpha = \sum_{l=0}^L w_l$, and an approximate implementation of $\hamiltonian$ with error $\epsilon_{R,H}$, $\hamiltonian^\prime = \sum_{l = 0}^L (w_l + \epsilon_{R,H}) \hamiltonian_l / \alpha^\prime$, $\alpha^\prime = \sum_{l=0}^L (w_l + \epsilon_{R,H})$.
Here $w_l$ are the LCU coefficients and $\hamiltonian_l$ are unitary operators.

Next, consider state preparation implemented using an identical set of phases $\{\phi_d\}$, but the truncated Hamiltonian $\hamiltonian^\prime$, under all conditions specified in Theorem 1. 
The final state produced by this process, $|\agsf^\prime\rangle$, can be written as
\begin{equation}
|\agsf^\prime\rangle = \overlap_f^\prime|\gs^\prime\rangle + \sqrt{1 - (\overlap_f^\prime)^2}|\Psi_{ex}^\prime\rangle,
\end{equation}
where $|\gs^\prime\rangle$ is the ground state of $\hamiltonian^\prime$ and $|\Psi_{ex}^\prime\rangle = |\agsi\rangle - \langle \agsi|\gs^\prime \rangle |\gs^\prime\rangle$ is the projection of the initial state $|\agsi\rangle$ onto the orthogonal complement of the subspace spanned by the ground state of $\hamiltonian^\prime$.
The overlap of $|\agsf^\prime\rangle$ with the exact ground state of $\hamiltonian$, $|\gs\rangle$, is then
\begin{equation}
    \overlap_f = \langle \agsf^\prime | \gs \rangle = \overlap_f^\prime\langle \gs|\gs^\prime \rangle + \sqrt{1 - (\overlap_f^\prime)^2}\langle \gs | \Psi_{ex}^\prime\rangle,
\end{equation}
which will generally be less than or equal to $\overlap_f^\prime$ in modulus.
Our goal is then to understand how $\overlap_f$ and $\overlap_f^\prime$ are related via $\epsilon_{R,H}$, which we can do by bounding the overlaps $\langle \gs|\gs^\prime \rangle$ and $\langle \gs | \Psi_{ex}^\prime\rangle$.

We can use perturbation theory to approximate $|\gs\rangle$ to lowest order in $\epsilon_{R,H}$ in the eigenbasis of $\hamiltonian^\prime$.
We first note that 
\begin{equation}
    \hamiltonian = \hamiltonian^\prime + \epsilon_{R,H} (\frac{L}{\lambda}\frac{\sum_i w_i\hamiltonian _i}{\lambda} - \sum_i \frac{\hamiltonian_i}{\lambda}) \equiv \hamiltonian^\prime + \epsilon_{R,H} \hamiltonian^{(1)}
\end{equation}
We can then construct the first order correction to $\gs$ using standard perturbation theory arguments, and back out the overlaps required.
Non-degenerate perturbation theory applies as the first excited state has a gap $\Delta$ from the ground state.

We start with the normalized (to $O(\epsilon_{R,H})$) perturbed state, denoted with a bar
\begin{equation}
    \bar{\gs} = \gs^\prime + \epsilon_{R,H}\sum_{k > 0} \frac{ \langle \Psi_k^\prime |\hamiltonian^{(1)}|\gs^\prime \rangle }{E_0^\prime - E_k^\prime} \Psi_k^\prime + O(\epsilon_{R,H}^2),
\end{equation}
and bound the second term on the right hand side by noting that $|
E_0^\prime - E_k^\prime | \geq \Delta/\alpha$ and that $ |\langle \Psi_k^\prime |\hamiltonian^{(1)}|\Psi_0^\prime \rangle| \leq L/\alpha$: 
\begin{equation}
    |\frac{ \langle \Psi_k^\prime |\hamiltonian^{(1)}|\Psi_0^\prime \rangle }{E_0^\prime - E_k^\prime}| \leq \frac{L}{\Delta}
\end{equation}
We can the compute bounds on the normalized overlaps as follows:
\begin{equation}
    \langle \gs | \gs^\prime \rangle = 1 + O(\epsilon_{R,H}^2), \ \
    |\langle \gs | \Psi_{ex}^\prime \rangle| \leq \epsilon_{R,H} \frac{L}{\Delta} + O(\epsilon_{R,H}^2).
\end{equation}

Plugging the computed bounds into our relationship between $\overlap_f$ and $\overlap_f^\prime$, we find a worst case lower bound
\begin{equation}
    \overlap_f \geq \overlap_f^\prime - \epsilon_R\sqrt{1 - {\overlap_f^\prime}^2}\frac{L}{\Delta} + O(\epsilon_{R,H}^2).
\end{equation}
Therefore, to ensure we can get an overlap $\overlap_f$ with a rotation synthesis error $\epsilon_{R,H}$, we would need to replace $\overlap_f$ with $\overlap_f + \epsilon_{R,H}\sqrt{1 - {\overlap_f}^2}\frac{L}{\Delta}$ in Eq~\ref{eq:final_epsilon}.

\subsection{Rotation synthesis error in \texorpdfstring{$e^{i\phi_j\Pi}$}{exp(iφⱼΠA)}}
The basis for our analysis is the expression for the QSP unitary $U_{\Phi}(x)$ in the qubitized basis, Eq~13 of Dong \textit{et al.}~\cite{dong2021efficient}
\begin{equation}
    U_{\Phi}(x) = e^{i\agsi\sigma_z}\Pi_{j=1}^{d}[W(x)e^{i\phi_j\sigma_z}],
\end{equation}
where $\Phi = \{\phi_j\}$ are the QSP phases, $d$ the degree of the QSP polynomial, and $W(x)$ the qubitized Hamiltonian operator, which is also unitary.
We can also write down the approximate QSP unitary due to a rotation synthesis error $U_{\Phi + \epsilon_{R, P}}$:
\begin{equation}
    U_{\Phi + \epsilon_{R, P}}(x) = e^{i(\agsi + \epsilon_{R, P})\sigma_z}\Pi_{j=1}^{d}[W(x)e^{i(\phi_j + \epsilon_{R, P})\sigma_z}].
\end{equation}
Our goal will be to compute a bound on the difference between the approximate and exact QSP unitary, and determine how the approximation affects the final overlap from QSP.

We begin by Taylor expanding the exponentials in $U_{\Phi + \epsilon_{R, P}}$ to first order in $\epsilon_{R, P}$, finding that each exponential gives a single correction term to the exact expression:
\begin{equation}
\begin{split}
U_{\Phi + \epsilon_{R, P}} - U_{\Phi} = &e^{i\agsi\sigma_z}i\epsilon_{R, P}\sigma_z\Pi_{j=1}^{d}[W(x)e^{i\phi_j\sigma_z}] + \\
&e^{i\agsi\sigma_z}\sum_{k=1}^{d}\Big(\Pi_{j=1}^{k-1} W(x)e^{i\phi_j\sigma_z)}\Big)i\epsilon_{R, P}\sigma_ze^{i\phi_k\sigma_z}\Big(\Pi_{j=k+1}^d W(x)e^{i\phi_j\sigma_z)}\Big) + O(\epsilon_{R, P}^2)
\end{split}
\end{equation}
Noting that each term on the RHS is a product of unitaries, we can compute a bound on the difference of QSP operators as such:
\begin{equation}
|U_{\Phi + \epsilon_{R, P}} - U_{\Phi}| \leq (d+1)\epsilon_{R, P} + O(\epsilon_{R, P}^2).
\end{equation}

We also know that both $U_{\Phi}, U_{\Phi + \epsilon_{R, P}}$ have identical eigenvectors, and that if all the conditions in Theorem 1 are met, the eigenvalues of $U_{\Phi}$ are $\epsilon$-close to $\pm 1$.
As such, if all conditions in Theorem 1 are met, the eigenvalues of $U_{\Phi + \epsilon_{R, P}}$ must be, in the worst case, $\epsilon + (d+1)\epsilon_{R, P}$-close to $\pm 1$.
Following the analysis in Appendix A, we can therefore replace $\epsilon$ with $\epsilon + (d+1)\epsilon_{R, P}$ in Eq~\ref{eq:final_epsilon} to account for the rotation synthesis error in the QSP phases.

The issue is, however, that the degree of the polynomial $d$ depends on $\epsilon$. 
As such, the replacement above cannot be carried about in such a trivial way, as we would like to isolate $\epsilon$.
In general the full analysis for isolating $\epsilon$ is very complex, but we can consider the analytics in a useful limit for $\epsilon, \delta \ll 1$ which we have already demonstrated is the region where QSP is a useful technique.
Under the limit $\epsilon, \delta \rightarrow 0$, we find that $d \rightarrow \frac{2}{\delta}\log{\frac{1}{\epsilon}}$ as demonstrated in Lin and Tong ~\cite{lin2020optimalpolynomial,lin2020near}.
Since we know that the error due to $\epsilon_{R, P}$ must be subleading, we can assert that to lowest order the correction to $\epsilon$ is linear and can be written as $\epsilon = \epsilon_0 + p$, where $\epsilon_0$ is the bare expression in Eq~\ref{eq:final_epsilon} and $|p|\ll1$ is the correction to the rotation synthesis error.

Replacing $\epsilon \rightarrow \epsilon + (d+1)\epsilon_{R, P}$ in Eq~\ref{eq:final_epsilon}, and applying our limits we find the following expression for $p$:
\begin{equation}
    \epsilon_0 + p  +\frac{2}{\delta}{\log\frac{1}{\epsilon_0 + p}}\epsilon_{R, P} \leq \epsilon_0.
\end{equation}
Taylor expanding the logarithm, and rearranging terms, we find that $p \leq \frac{-2}{\delta} \log{\frac{1}{\epsilon_0}}\epsilon_{R, P} / (1 - 2\epsilon_{R, P}/\epsilon_0\delta)$.
Noting that the rotation synthesis error is only subleading if $(d+1)\epsilon_{R, P} \ll \epsilon_0 \rightarrow \frac{2\epsilon_{R, P}}{\delta} \frac{\log{\frac{1}{\epsilon_0}}}{\epsilon_0} \ll 1 \rightarrow 2\epsilon_{R, P}/\epsilon_0\delta \ll 1$, we find a final expression for $p$ and thereby $\epsilon$:
\begin{equation}
    p \leq -\frac{2}{\delta}\epsilon_{R, P} \log{\frac{1}{\epsilon_0}},\ \epsilon = \epsilon_0 + p
\end{equation}

\subsection{Net effect of rotation synthesis error}
We combine the rotation synthesis errors from the previous two sections into a correction to the bound on $\epsilon$ in Theorem 1,
\begin{equation}
    \epsilon \leq \frac{1 - \overlap_f^2}{6N_{iter}^2} -
   \frac{|\overlap_f|\sqrt{1 - \overlap_f^2}}{4N_{iter}^2}\frac{L}{\Delta}\epsilon_{R, H} - 
    \frac{2}{\delta}\log{\Big(\frac{1 - {\overlap_f^2}}{6N_{iter}^2}\Big)^{-1}}\epsilon_{R, P} + O(\epsilon_{R, P}^2).
    \label{eq:rotsynth_epsilon}
\end{equation}

To understand this expression, we can first look at the sign of $\epsilon$.
We need to ensure that $\epsilon$ is bounded above by a positive number, otherwise we cannot accomplish the task of boosting to overlap $\overlap_f$ at all.
In order to do this, we will require that the two correction terms are much smaller than the first base term.
We note that $N_{iter} \sim |\overlap_i|^{-1} \sim e^{\eta/2}$ where $\eta$ is the system size and $\Delta$ is a constant as $
\eta\rightarrow \infty$.
To ensure that the second term on the RHS is smaller than the first, we will require $\epsilon_{R, H} \sim 1/L$, which follows the same scaling with system size determined by Babbush \textit{et al.} for QPE~\cite{babbush2018encoding}.
Since rotation synthesis errors only appear logarithmically in the T counts, prefactors become negligible and we use the rotation synthesis bounds from Babbush's work for our Hamiltonian implementation.

To ensure that the third term on the RHS is smaller than the first, we require $\epsilon_{R, P} \sim e^{-\eta}/(\eta N)$.
While this seems concerning at first thought, we note that rotation synthesis appears in the T counts in a logarithmic factor, ensuring that rotation operators have depth $\log\left(e^{-\eta}/(\eta N)\right) \in \tilde{O}(\eta)$ for large system sizes.
Additionally, the high-depth rotations are only required for the QSP phases, which are applied once per application of an entire Hamiltonian block-encoding.
We include high-depth rotations for the QSP single-qubit rotation implementations using the value $\epsilon_{R, P} = 10^{-10} \epsilon_0/d$. 

\section{Appendix C: Computing \texorpdfstring{$\mu$}{μ} and \texorpdfstring{$\Delta$}{Δ}}
\setcounter{equation}{0}
\setcounter{figure}{0}
\renewcommand{\theequation}{C\arabic{equation}}
\renewcommand{\thefigure}{C\arabic{figure}}
\label{app:computing_gap_and_bound}
We require bounds on two parameters, $\mu$ and $\Delta$, to implement the approximate reflector $\mathcal{R}_{\gs}(\epsilon)$.
Specifically, we require the bounds to satisfy the constraints $E_0 \leq \mu - \Delta /2 < \mu + \Delta/2 \leq E_1$.
Generically, finding parameters $\mu$ and $\Delta$ is difficult, as they are related to the ground- and first-excited state energies of $\hamiltonian$.
In this Appendix we describe the methods we have used to find values for the parameters $\mu$ and $\Delta$ that satisfy the above bounds.

For the TFIM calculations we have analytic results for the spectrum which help us compute the values of $\mu$ and $\Delta.$
It is known that the gap in the TFIM defined in Eq.~\ref{eq:tfim} is $2(|g| - 1)$ when $|g| > 1$ and $2(1 - |g|)$ for $|g| < 1$ \cite{tfimreview}.
We ignore the case of $|g| = 1$ as the TFIM is gapless, and the state preparation considered in this work requires a spectral gap to be present.
We can compute the exact ground state energy $E_0$ for the TFIM by transforming the spin operators in the Hamiltonian using the Jordan-Wigner transformation, which results in a non-interacting Hamiltonian that is easily diagonalized on a classical computer.
Therefore, knowing both $E_0$ and $E_1$ exactly for the TFIM, we can simply set $\mu = (E_1 + E_0)/2$ and $\Delta = (E_1 - E_0)$.

For the first quantized calculations, we use the binary search algorithm of Lin and Tong \cite{lin2020near} to compute an approximation to the ground state energy $E_0$ to an accuracy of $\delta_1$: $E_0^{\text{bin}} = E_0 + \delta_1$.
To reduce the overhead from conducting the binary search over the range $[-\alpha, \alpha]\ E_h$, we reduce the binary search range to $[\bar{E}_0 - N_{atom}/2, \bar{E}_0 + N_{atom}/2]\ eV$ where $\bar{E}_0$ is an approximate ground state energy computed from a classical simulation which we take to contain the exact ground state energy within 1 eV/atom error.
One can search over a larger range at the cost of a logarithmic expense in the range.
We note that for a 1 eV/atom range, the logarithmic prefactor for a typical first quantized calculation is $\sim$5, while the barebones search over $\pm \alpha$ has a logarithmic prefactor $\sim$50.
This prefactor can be further reduced by implementing the requisite phase estimation with an adaptive variance, i.e., estimating with a larger $\Delta E$ for the larger intervals.
We find an approximation to the true gap $E_1 - E_0$ by looking at the experimentally measured value for the gap $\Delta^{\text{exp}}$.
In general this will contain an error such that $\Delta^{\text{exp}} = (E_1 - E_0) + \delta_2$.

Based on our knowledge of the magnitudes of $\delta_1$ and $\delta_2$, we compute values of $\mu$ and $\Delta$ that satisfy the required bounds.
We assume that $\delta_2 \leq (E_1 - E_0)$ so that the experimentally measured gap is within at most 100\% away of the exact gap.
With this assumption, we choose $\delta_1 = \Delta^{\text{exp}}/6$, and set $\mu = \Delta^{\text{exp}}/6 + E_0^{\text{bin}}$ and $\Delta = \Delta^{\text{exp}}/3$.
Double checking, we find that $\mu + \Delta/2 \leq \Delta^{\text{exp}}/2 + E_0 \leq E_1$, satisfying the upper bound.
A similar calculation for the lower bound yields $\mu - \Delta/2 = E_0 + \Delta^{\text{exp}}/2 \geq E_0$ as required.
One can adjust all of the scaling factors appropriately based on the knowledge of $\delta_2$, but it should be noted that reducing $\Delta$ increases the cost the most due to the $\Delta^{-1}$ scaling of AA. 
For the calculation on the $\beta$-aluminas in this work we use $\Delta^{\text{exp}}$ = 9 eV \cite{aluminas_battery}.

\section{Appendix D: Alternative definitions of \texorpdfstring{$\improvement$}{ι}}
\setcounter{equation}{0}
\setcounter{figure}{0}
\renewcommand{\theequation}{D\arabic{equation}}
\renewcommand{\thefigure}{D\arabic{figure}}
\label{app:alternative_improvement}

$\improvement$ is intended to quantify the improvement in the expected runtime associated with implementing AA in ground state energy estimation, relative to not implementing it and repeating QPE over more trials.
Short of mapping either algorithm onto a more detailed fault-tolerant quantum computer architecture, this ratio of runtimes is approximated by the ratio of $T$ counts.
In what follows we briefly consider further refinements of this ratio that might more accurately capture the expected speedup.

In a model of fault-tolerant quantum computation based on the surface code, the runtime of either algorithm will be proportional to the distance of the logical qubits used in their implementation.
We denote the distance with (without) AA as $d_{AA}$ ($d_{\overline{AA}}$).
If the two algorithms are implemented with the same distance, then these factors will cancel in a runtime ratio, consistent with the definition of $\iota$ in the main text.
If the two algorithms use different distances then $\iota$ should be scaled by the proportionality factor $d_{\overline{AA}}/d_{AA}$.

A scenario in which it might be useful to consider running the two algorithms at different distances is one in which having a machine large enough to implement the algorithm with AA naturally enables the implementation of the algorithm without AA at a higher distance.
This would improve the probability of successfully implementing any given run of QPE by further reducing the logical error rate of any given logical qubit.
In this scenario, $d_{\overline{AA}}/d_{AA} > 1$ because the runs without AA will have a longer clock cycle, and making the run without AA slower will naturally increase the speedup for the AA algorithm.
However, there is also a scenario in which $d_{\overline{AA}}/d_{AA} < 1$, reducing the speedup for the AA algorithm.
For a scenario in which the probability of successfully implementing the algorithm with AA is the same or greater than the probability of successfully implementing the algorithm without AA, the logical error rate for any given operation will need to be lower for the former.
Assuming that the success probabilities are the same for the two algorithms, $d_{\overline{AA}}/d_{AA}$ is asymptotically logarithmic in the quadratic speedup factor in Eq.~\ref{eq:asymptotic_improvement}, i.e., it scales inversely with $\log\left(\frac{\Delta}{\Delta E} \overlap_i\right)$.

\section{Appendix E: Background}
\setcounter{equation}{0}
\setcounter{figure}{0}
\renewcommand{\theequation}{E.1.\arabic{equation}}
\renewcommand{\thefigure}{E.1.\arabic{figure}}
\subsection{Block Encoding}
The Hamiltonian itself is not unitary and therefore not amenable to direct application on a quantum computer;
however, the Hamiltonian can be embedded into a larger Hilbert space such that the entire operation \emph{is} unitary through a process known as ``block encoding" \cite{gilyen2019quantum}.
One common strategy for block encoding is the linear combination of unitaries (LCU) approach \cite{childs2012lcu} where the Hamiltonian is written as
\begin{equation}
    \hamiltonian = \sum_{l=0}^{L} \norm{l} H_l.
    \label{eq:h_lcu}
\end{equation}
Each $H_l$ is a $n$-qubit unitary acting on the system register, and the number of terms $L$ necessarily sets the size of the $m$-qubit auxiliary register such that $L < 2^m - 1$.

We can then define a pair of $m$-qubit state preparation unitaries
\begin{equation}
\begin{split}
    \prep{H} \ket{0}^{\otimes m} = \sum_{l=0}^L a_l \ket{l}, \\
    \unprep{H} \ket{0}^{\otimes m} = \sum_{l=0}^L b_l \ket{l},
    \label{eq:lcu_prep_unprep}
\end{split}
\end{equation}
with coefficients satisfying
\begin{equation}
    \sum_{l=0}^{L} \left\vert a_l^* b_l - \frac{\norm{l}}{\norm{}}\right\vert < \epsilon_{\rm LCU},
    \label{eqn:spp_condition}
\end{equation}
where $\norm{} = \Vert \norm{l} \Vert_1$ is the norm and $\epsilon_{\rm LCU}$ is the error of the Hamiltonian LCU block encoding.
In the case where $\norm{l} > 0$, one can set $a_l = b_l = \sqrt{\norm{l}}$ and the state preparation pair reduces to a single prepare oracle.
Even in the case of negative coefficients, the two oracles often only differ by the sign on several rotation angles.

We can also define the $(n+m)$-qubit select unitary as
\begin{equation}
    \sel{H} = \sum_{l=0}^{L} \vert l\rangle\langle l\vert \otimes H_l,
    \label{eq:lcu_sel}
\end{equation}
which applies each $H_l$ conditioned on the value of the auxiliary register.
The full block encoded Hamiltonian is then given by
\begin{equation}
    U_H = (\unprep{H}^\dagger \otimes I_n) \sel{H} (\prep{H} \otimes I_n),
    \label{eq:h_blockencoded}
\end{equation}
which is the core operation used in both quantum phase estimation and ground state preparation.
For brevity and due to their often shared implementation, $\unprep{}^\dagger$ may simply be denoted $\prep{}^\dagger$ in the remainder of the paper.

Therefore, determining efficient circuit implementations and resource estimates for the PREP, UNPREP, and SEL oracles is the critical task for resource estimation of the entire algorithm.

\subsection{Circuits, T counts and qubits for transverse field Ising Model (TFIM)}
\setcounter{equation}{0}
\setcounter{figure}{0}
\renewcommand{\theequation}{E.2.\arabic{equation}}
\renewcommand{\thefigure}{E.2.\arabic{figure}}
\label{app:tfim_circuits}
\subsubsection{Hamiltonian}
\label{app:tfim_hamiltonian}
The LCU implementation of the $\hamiltonian_{TFIM} + \mu I$, with $\hamiltonian_{TFIM}$ defined as
\begin{equation}
\hamiltonian_{TFIM} = \sum_{i =1}^L S_i^z S_{i+1}^z + g S_i^x 
    \label{eq:tfim}
\end{equation}
requires the specification of three circuit components: $\prep{TFIM}, \unprep{TFIM}$ and $\sel{TFIM}$.
As $\mu < 0$ is in general the two circuits $\prep{TFIM}$ and $\unprep{TFIM}$ will not be identical.

We begin with the $\prep{TFIM}$ and $\unprep{TFIM}$ circuits.
Following the prescription in Eq~\ref{eq:lcu_prep_unprep}, we will require $m = \lceil log_2 L \rceil + 2$ auxiliary qubits for an $L$ site TFIM.
To understand the additional two qubits, we need to count the total number of coefficients in the system.
There are $2L$ coefficients for each term in $\hamiltonian_{TFIM}$ and a single additional coefficient $\mu$.
In order to implement the first set of $2L$ coefficients we require $\lceil \log_2 L \rceil + 1$ qubits in register $|m\rangle$, and the additional coefficient $\mu$ requires one more qubit.
We will refer to the registers then as such: $|j\rangle$ which has size $\lceil \log_2 L \rceil $ and two single-qubit registers $|g\rangle$ and $|\mu\rangle$ which implement the unique coefficients $g, \mu$.

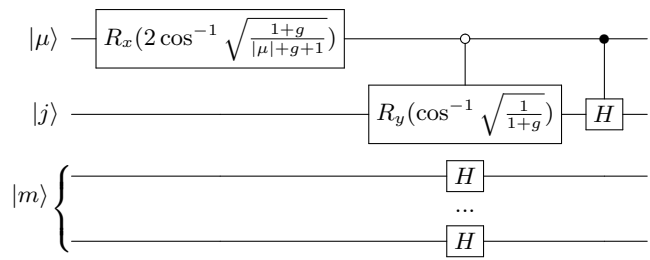
\begin{figure}[h]
\mbox{
\Qcircuit @C=1em @R=.7em 
{\lstick{\ket {\mu}}& \gate{R_x(2 \cos^{-1}\sqrt{\frac{1+g}{|\mu| + g + 1}})} & \ctrlo{1} & \ctrl{1} & \qw \\
\lstick{\ket{j}}& \qw & \gate{R_y( \cos^{-1} \sqrt{\frac{1}{1+g}})} & \gate{H} & \qw \\
& \qw & \gate{H} & \qw & \qw \\
& & ... & &  \\
& \qw & \gate{H} & \qw & \qw \inputgroupv{3}{5}{.8em}{.8em}{\ket{m}} \\
}
}
\caption{Implementation of $\prep{TFIM}$ over three registers: $|\mu\rangle, |j\rangle$ with a single qubit each and $|m\rangle$ with $\lceil \log_2 L \rceil$ qubits.
}
\label{fig:tfim_prep}
\end{figure}

The full circuit for $\prep{TFIM}$ is presented in Fig~\ref{fig:tfim_prep}, and can be easily shown to satisfy Eq~\ref{eq:lcu_prep_unprep}.
Following that is the $\unprep{TFIM}$ in Fig~\ref{fig:tfim_unprep} has a slight modification due to the $\mu < 0$ coefficient, namely that we have a minus sign on the rotation angle for the $R_x$ gate on the $|\mu\rangle$ register.

\begin{figure}[h]
\mbox{
\Qcircuit @C=1em @R=.7em 
{\lstick{\ket {\mu}}& \gate{R_x(-2 \cos^{-1}\sqrt{\frac{1+g}{|\mu| + g + 1}})} & \ctrlo{1} & \ctrl{1} & \qw \\
\lstick{\ket{j}}& \qw & \gate{R_y( \cos^{-1} \sqrt{\frac{1}{1+g}})} & \gate{H} & \qw \\
& \qw & \gate{H} & \qw & \qw \\
& & ... & &  \\
& \qw & \gate{H} & \qw & \qw \inputgroupv{3}{5}{.8em}{.8em}{\ket{m}} \\
}
}
\caption{Implementation of $\unprep{TFIM}$ over three registers: $|\mu\rangle, |j\rangle$ with a single qubit each and $|m\rangle$ with $\lceil \log_2 L \rceil$ qubits. \stefan{I think we probably could just say "For $\unprep{TFIM}$, the angle in the $R_x$ gate is negated" in Fig. \ref{fig:tfim_prep} or something. The difference is subtle enough that I'm not sure readers will easily notice. Also, I think that it still works if they are $R_y$ instead of $R_x$ - the angle negation is the important part. Not an action item, just FYI.}
}
\label{fig:tfim_unprep}
\end{figure}

\begin{figure}[h]
\mbox{
\Qcircuit @C=1em @R=1.5em 
{\lstick{\ket {\mu}}& \ctrlo{1} & \ctrlo{1} & \ctrlo{1} & \ctrlo{1} & \ctrlo{1} & \ctrlo{1} & \ctrlo{1} & \ctrlo{1} & \qw \\
\lstick{\ket{j}}& \ctrlo{1} & \ctrlo{1} & \ctrlo{1} & \ctrlo{1} & \ctrl{1} & \ctrl{1} & \ctrl{1} & \ctrl{1} & \qw \\
& \ctrlo{1} & \ctrlo{1} & \ctrl{1} & \ctrl{1} & \ctrlo{1} & \ctrlo{1} & \ctrl{1} & \ctrl{1} & \qw\\
& \ctrlo{1} & \ctrl{2} & \ctrlo{3} & \ctrl{1} & \ctrlo{1} & \ctrl{2} & \ctrlo{3} & \ctrl{4} & \qw \inputgroupv{3}{4}{.8em}{.8em}{\ket{m}} \\
& \gate{Z} \qwx[1] & \qw & \qw & \gate{Z} \qwx[3] & \gate{X} & \qw & \qw & \qw & \qw \\
& \gate{Z} & \gate{Z} \qwx[1] & \qw & \qw & \qw & \gate{X} & \qw & \qw & \qw \\
& \qw & \gate{Z} & \gate{Z} \qwx[1] & \qw  & \qw & \qw & \gate{X} & \qw & \qw \\
& \qw & \qw & \gate{Z} & \gate{Z} & \qw & \qw & \qw & \gate{X} & \qw 
\inputgroupv{5}{8}{.8em}{.8em}{\ket{\psi}} \\
}
}
\caption{Implementation of $\sel{TFIM}$ over three registers: $|\mu\rangle, |j\rangle$ with a single qubit each, $|m\rangle$ with $\lceil \log_2 L \rceil$ qubits, and $|\psi\rangle$ the state register with $L$ qubits.
}
\label{fig:tfim_sel}
\end{figure}
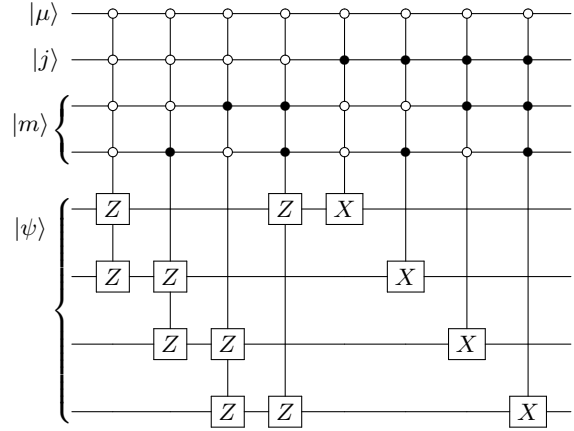

The circuit $\sel{TFIM}$ is slightly more complex, due to the need for controlled gates.
An implementation of $\sel{TFIM}$ for $L=4$ is presented in Fig~\ref{fig:tfim_sel}, which has has generic features which allow for easy translation to larger $L$.
The first obvious difference from the $\prep{TFIM}, \unprep{TFIM}$ circuits is the presence of a state register $|\psi\rangle$ with size $L$ that stores the wave function for the $L$ site TFIM. 
The only gates that need to be applied to the register $|\psi\rangle$ are $Z$ and $X$ gates which are controlled by the registers $|\mu\rangle$, $|j\rangle$ and $|m\rangle$.
The $Z$ gates are only applied with $|\mu\rangle = |0\rangle$, $|j\rangle = |0\rangle$, and $X$ gates only when $|\mu\rangle = |0\rangle$, $|j\rangle = |1\rangle$.
Two $Z$ gates are applied and one $X$ for each possible configuration of the register $|m\rangle$, with each of the $L$ possible configurations corresponding to the site $i$ which the operator $\overlap_i^{z}S_{i+1}^Z$ or $\overlap_i^x$ is applied.
In general then, we have $2L$ applications of a controlled $Z$ and $L$ applications of a controlled $X$ over the combined register $|\mu, j, m\rangle$.

\subsubsection{T counts: Hamiltonian}
The T counts are easily read-off the diagrams.
Since we have written everything in terms of single-qubit rotations gates, we use the compilation model of $\lceil10 + 4\log2_ \frac{1}{\epsilon_{PREP}}\rceil$ $T$ gates per rotation to compute T counts \cite{selingerR}.
It should be noted that the controlled rotations can be decomposed as four single-qubit rotations and two CNOTs, with the latter not requiring any $T$ gates to implement
The full $T$ counts for the $\prep{TFIM}$ and $\unprep{TFIM}$ is 
\begin{equation}
T[\prep{TFIM}] + T[\unprep{TFIM}] = 10(10 + 4\log_2{\frac{1}{\epsilon_{PREP}}})
\label{eq:tfim_prep_unprep_t}.
\end{equation}
We can similarly compute the T counts for the $\sel{TFIM}$ circuit, which has no rotations but many multi-controlled gates.
Reading off the T counts for multi-controlled X gates using the compilation model in \cite{niemannMCX}
\begin{equation}
    T[\sel{TFIM}] = 3L \times 24(\lceil \log_2 L \rceil + 2).
\label{eq:tfim_sel_t}
\end{equation}
It should be noted that using a sawtooth or unary implementation from \cite{babbush2018encoding} for the control logic can reduce the T counts for the $\sel{TFIM}$ circuit, but given how lightweight the circuits for TFIM already, sophisticated logic implementations were not included in our implementation.

The full T counts for implementing the $L$ site TFIM Hamiltonian with a shift $\mu$ is then:
\begin{equation}
T[\mathcal{U}({\hamiltonian_{TFIM}})] = 100 + 40\log_2 \frac{1}{\epsilon_{PREP}} + 6L + 72L \lceil \log_2 L \rceil
\label{eq:tfim_hamiltonian_t}
\end{equation}

\subsubsection{Logical qubits: Hamiltonian}
For the second quantized calculation, the total number of qubits  required for state preparation are simply those required for block-encoding of the Hamiltonian, plus an additional single QSP qubit.
As mentioned above, the $|\mu\rangle$ register requires a single qubit, the $|j\rangle$ register a single qubit, and $|m\rangle$ requiring $\lceil \log_2L \rceil$ qubits.
As the state register $|\mu\rangle$ requires $L$ qubits, we have a total qubit count for implementing the Hamiltonian of
\begin{equation}
    N_{\text{qubits}} = 2 + \lceil \log_2 L \rceil + L.
\end{equation}

\subsection{Circuits, T counts and qubits for first quantized systems}
\setcounter{equation}{0}
\setcounter{figure}{0}
\renewcommand{\theequation}{E.3.\arabic{equation}}
\renewcommand{\thefigure}{E.3.\arabic{figure}}
\label{app:first_quant_circuits}
\subsubsection{Antisymmetrization}
\label{app:implementation-antisym}
Here, we briefly describe the anti-symmetrization procedure of \cite{berry2018improved}, as employed in our resource estimates.
The overall idea is to prepare an antisymmetrization of $\ket{0}$\ldots $\ket{\eta-1}$; a later step will perform a transformation $\ket{i}\mapsto \ket{\varphi_i}$.
To wit, the permutations of the $\eta$ system registers are represented as quantum states of a \texttt{record} register.
At a high level, a superposition of these permutations is produced, and then the system registers are (coherently) shuffled according to that permutation.
While performing that shuffling, the \texttt{record} register is simultaneously ``unwound'' so that the \texttt{record} and system registers are disentangled.
Concretely, the \texttt{record} register is a list of the results of all comparisons made during the execution of a sorting \emph{network}, which is a sorting algorithm in which predetermined comparisons are made between particular elements of the list, and swaps are performed to put them in the correct order.
The state of the sorting network, then, is fully determined by the order of the list being sorted, and the progress through the algorithm.
In particular, as suggested in \cite{berry2018improved}, we use the bitonic sorting network, which can also be parallelized (though we do not explicitly track circuit depth).

Producing the superposition of permutation in \texttt{record} is done by generating all strings of length $\eta$, and filtering them down to bonafide permutations of the system register.
The filtering step is performed by looking for any duplicates in the string, and discarding them.

The only difference from \cite{berry2018improved} is that we separate the filtering stage, which generates \texttt{record}, from the shuffling stage, which actually antisymmetrizes the system register.
The shuffling step is unitary, and must be repeated for every application of $\mathcal{R}_{\agsi}$, while the filtering step need only be performed once per shot~\footnote{Actually, the filtering step succeeds with probability $>1/2$, so on average the step needs to be performed twice.}.
\vspace{1em}
%

\paragraph{Startup/filtering}
\begin{enumerate}
\item Prepare $(\ket{0}+\ket{1})^{\otimes (\eta \lceil\log_2 \eta^2\rceil)}$.
Interpret this as a superposition of length-$\eta$ strings over an alphabet of $\eta^2$ characters.
\item Sort the string using a bitonic sorting network.  Store each comparison result in \texttt{record}.
\item Detect repeated terms in the (now sorted) string; if found, restart the algorithm.
\end{enumerate}
A key result of \cite{berry2018improved} is that repeated terms happen with probability $<1/2$, provided the initial string is over a large enough alphabet---at least the square of $\eta$, as we have here.
Additionally, the \texttt{record} register will be disentangled, and in a superposition of all sorted repetition-free strings of length $\eta$.
\vspace{1em}

\paragraph{Shuffling/permuting}
Run the bitonic sorting network again, with the following performed at each step:
\begin{enumerate}
\item At every step of the sorting algorithm network, swap system registers if the \texttt{record} comparison indicates a swap was performed.
\item If a swap was performed in the above step, act a $Z$ gate on the same qubit of \texttt{record}.  This adds a $-1$ phase for every swap performed, and therefore produces an antisymmetrized wavefunction.
\item Compare the values of the (possibly) swapped system registers.  Use this to zero the value of the qubit in \texttt{record}.
\end{enumerate}
Notice that the above shuffling step leaves \texttt{record} disentangled from the system register, though it requires that the system registers start out in a sorted state.
Also, not that the process can be reversed: one can act this circuit backwards to take an antisymmetrized state back to a prepared \texttt{record} state, and a sorted system register.

\subsubsection{T counts: Antisymmetrization}
The bitonic sort network has a deterministic progression of comparison/swap steps.
The number of those steps is bounded above by
\begin{equation}
C = \left\lfloor\frac{\eta}{2}\right\rfloor\frac{\lfloor\log_2\eta\rfloor(\lfloor\log_2\eta\rfloor+1)}{2}
\end{equation}

Following \cite{berry2018improved}, the $T$ cost of comparisons is $8\lceil\log_2\eta^2\rceil$, and each controlled-swap can be performed at a $T$ cost of $4\lceil\log_2\eta^2\rceil$ \aer{to get this 4, you need to do 1 measurement per controlled-swap, and classical feedforward; if that's not ok, you can do this by catalytically using $1$ $T$ state (i.e., it's recovered afterwards), and $6$ $T$ gates}.
This leads to a total cost of the startup circuit:
\begin{equation}
    T_\text{antisym-startup} = \left\lfloor\frac{\eta}{2}\right\rfloor\frac{\lfloor\log_2\eta\rfloor(\lfloor\log_2\eta\rfloor+1)}{2}\left(4\lceil\log_2\eta^2\rceil+8\lceil\log_2\eta^2\rceil\right)+8(\eta-1)\lceil\log_2\eta^2\rceil.
\end{equation}

The shuffle step does not need the test for non-repetition, but is otherwise the same cost:

\begin{equation}
    T_\text{antisym-shuffle} = \left\lfloor\frac{\eta}{2}\right\rfloor\frac{\lfloor\log_2\eta\rfloor(\lfloor\log_2\eta\rfloor+1)}{2}\left(4\lceil\log_2\eta^2\rceil+8\lceil\log_2\eta^2\rceil\right).
\end{equation}

\subsubsection{Qubit counts: Antisymmetrization}
There are $\eta \lceil \log_2 N \rceil$ qubits used in the system register, and $\eta \lceil \log_2 \eta^2 \rceil$ qubits used to construct the \texttt{record}.
The comparison circuit has an overhead of two qubits, so an additional $2\log_2\eta$ qubits is needed during the sorting and shuffling stages.
Similarly, the Fredkin (controlled-swap gate) have an overhead of one qubit.
The total overhead for any part of the antisymmetrization stage running is then (at most) $3\log_2 N$.
Finally, the \texttt{record} of size $C =    \left\lfloor\frac{\eta}{2}\right\rfloor\frac{\lfloor\log_2\eta\rfloor(\lfloor\log_2\eta\rfloor+1)}{2}$ qubits.
The additional number of qubits required aside from the system register, with no additional handling, is then
\begin{equation}
    \eta \lceil \log_2 \eta^2 \rceil +
    3\lceil \log_2 \eta \rceil + 
    \left\lfloor\frac{\eta}{2}\right\rfloor\frac{\lfloor\log_2\eta\rfloor(\lfloor\log_2\eta\rfloor+1)}{2}.
    \label{eq:qubits_as}
\end{equation}
The qubits represented in the first term of Eq~\ref{eq:qubits_as} must be kept during the entire runtime, while the second and third term represent qubits that can be reused, for example in quantum phase estimation.

During anti-symmetrization we only need $\eta \lceil \log_2 \eta^2 \rceil$ qubits in the system register, meaning that $\eta \lceil \log_2 N - \log_2 \eta \rceil$ qubits are freed for use on the largest qubit overhead in Eq~\ref{eq:qubits_as}, namely \texttt{record}.
The total number of additional qubits other than the system register needed for anti-symmetrization can then be reduced to:
\begin{equation}
    \eta \lceil \log_2 \eta^2 \rceil +
    3\lceil \log_2 \eta \rceil + 
    \left\lfloor\frac{\eta}{2}\right\rfloor\frac{\lfloor\log_2\eta\rfloor(\lfloor\log_2\eta\rfloor+1)}{2} -
    \eta \lceil \log_2 N - \log_2 \eta \rceil.
\end{equation}
Due to the large qubit overhead in producing and using \texttt{record}, we can further reduce the total qubit count for state preparation and QPE by reusing \texttt{record} qubits when not running anti-symmetrization.

\subsubsection{Hartree-Fock
\label{app:implementation-rotmat}
}

Next, we demonstrate how to perform the unitary $V=\ket{j}\mapsto\ket{\varphi_j}$ on all of the system registers.
The most direct approach would simply be to find a circuit for $V$, and act it on each register.
However, as shown in \cite{delgado2022Apr}, there are significant advantages in exploiting the fact that the system register only contains each $\ket{j}$ once.
We briefly reproduce the algorithm here for completeness.

The first observation is that the unitary $V$,
\begin{equation}
V\ket{j} = \ket{\varphi_j},
\end{equation}
has a QR-decomposition using Givens rotations,
\begin{equation}
V = \prod_p \prod_{q>p} G_{pq} R,
\end{equation}
where $R$ is diagonal and $G_{pq}$ is unitary; both act non-trivially only on the $\ket{p}$-$\ket{q}$ subspace.
Moreover, because $V$ is unitary, $R$ is also unitary.
And, because the phases only contribute an irrelevant global phase, we ignore $R$ hereafter.

Next, we observe that we can truncate the expansion to
\begin{equation}
V = \prod_{p<\eta} \prod_{q>p} G_{pq},
\end{equation}
because we do not care about the behavior of $V$ on initial input states besides $\ket{0}\ldots\ket{\eta-1}$.
We can also truncate the behavior on the output states:
\begin{equation}
V = \prod_{p<\eta} \prod_{q\geq\eta} G_{pq}.
\end{equation}
Those additional terms would not affect the output state because any rotation of an occupied state into an occupied state will be a no-op, and we are only concerned with the behavior of $V$ on the antisymmetrized state $\mathcal{A} \ket{0}\ldots\ket{\eta-1}$.
Finally, notice that each $G_{pq}$ can be decomposed into a $Z$, $Y$ and $Z$ rotation (as well as irrelevant global phase).
Notice that initial (rightmost) $Z$ rotation can be absorbed into the diagonal $R$ and ignored.
This means that each $G_{pq}$ can be implemented (sufficient for our purproses) with two rotations.

Thus simplified, we now describe the implementation of $V^{\otimes \eta}$, i.e., the application of $V$ simultaneously on all $\eta$ system registers.
It turns out that the dominant cost of this broadcasted-$V$ operation is, for each Givens rotation $G_{pq}$, simply checking if the input state is in the state $\ket{p}$ or $\ket{q}$.
Once that has been determined, we can shuffle the affected register to a known location, and then conditionally apply the Givens rotation.
Concretely, the algorithm for performing the broadcasted Givens rotation, $G_{pq}^{\otimes \eta}$, is \cite{delgado2022Apr}:
\begin{enumerate}
\item Initialize $\eta$ auxiliary qubits.
\item For each register, $j$, detect if it needs to be rotated and store in the $j$th auxiliary qubit.
I.e., do two equality tests on register $j$ comparing to the integers $p$ and $q$, each controlling a flip on auxiliary register $j$.
\item For each register, $j$, swap the register with the final register controlled on the auxiliary register.
\item Notice that at most two auxiliary qubits can be flipped, because at most one of $p$ and $q$ each can be present in the antisymmetrized wavefunction.
Controlled on exactly one (equivalently, an odd number) of the auxiliary registers being set:
\begin{enumerate}
 \item Identify the first digit that differs in the binary representation of $p$ and $q$.  Controlled on that bit, flip the other digits that differ.
 Assuming the first difference in the binary representation is a bit that is $0$ in $p$, then, after doing this, the register will be $p$ (if it started in $p$), or $p$ with that digit flipped.
 \item Perform $Z$ and $Y$ rotations on that first differing binary digit as required to implement $G_{pq}$.
 \item Undo (equivalently, redo) step (4a).
\end{enumerate}
\item Undo steps 3 and 2.
\end{enumerate}

\subsubsection{T counts: Hartree-Fock}
The equality test in step $2$ takes a $\lceil\log_2 N\rceil$-Toffoli, which we assume has $T$ cost $24\lceil\log_2 N\rceil$.
Again, following \cite{berry2018improved}, each controlled-swap can be performed at a $T$ cost of $4\lceil\log_2\eta^2\rceil$\aer{to get this 4, you need to do 1 measurement per controlled-swap, and classical feedforward; if that's not ok, you can do this by catalytically using $1$ $T$ state (i.e., it's recovered afterwards), and $6$ $T$ gates}.
Additionally, we take the $T$ cost of rotations to be $\lceil 10-4\log_2\epsilon\rceil$.

\begin{equation}
 T_\text{Givens} = 2(2\cdot 24\lceil\log_2 N\rceil\cdot\eta + 4\cdot \lceil\log_2 N\rceil (\eta-1)) + 2\lceil 10-4\log_2(\epsilon_{R,\text{H-F}})\rceil
\end{equation}
where $\epsilon$ is the angular error permitted in the Givens rotations realizing the Hartree-Fock state.\aer{I feel like I am missing the rotation error notation we're using.  This needs to be brought in line.}\stefan{I have a similar out-of-line error notation for the first-quant Hamiltonian. I end up relating it back to the total error $\epsilon$ once, and otherwise don't worry about it too much}
The total cost for the Hartree-Fock step, then, is
\begin{equation}
 T_\text{HF} = \eta(N-\eta)T_\text{Givens}+T_\text{antisymm-shuffle}
\end{equation}

\subsubsection{Qubit Counts: Hartree-Fock}
An overhead of $\eta$ auxiliary qubits is required to track if the state is $p$ or $q$.
We perform the equality test with a Toffoli, so it requires only 1 additional qubit.
The controlled swap requires $\log_2N$ qubits.
These temporary qubit overheads sum to
\begin{equation}
    \eta+\log_2N+1.
\end{equation}

\subsubsection{Hamiltonian}

The electronic structure Hamiltonian under the Born-Oppenheiemer approximation, i.e. with fixed nuclear degrees of freedom, is generally given as
\begin{equation}
    \hamiltonian = T + U + V,
    \label{eq:h_tuv}
\end{equation}
where T is the electronic kinetic energy, U is the nuclear-electron attraction, and V is the electron-electron repulsion.
Constant terms such as nuclear-nuclear interactions are also present, but can be computed classically and therefore not included here.
Further specification of the Hamiltonian requires a choice of basis set;
for materials, one common choice is to use a plane wave basis in a unit cell with periodic boundary conditions.
Each plane wave is given by the equation
\begin{equation}
    \phi_p(r) = \sqrt{1/\Omega}e^{-i\,k_p\cdot r},
\end{equation}
where $r\,(k_p)$ is a vector in real (reciprocal) space and $\Omega$ is the volume of the unit cell.
As in Su \textit{et al.} \cite{su2021fault}, we will only consider a cubic reciprocal lattice of $N$ plane wave functions appropriate for system with no or cubic periodicity such that
\begin{equation}
    k_p = \frac{2\pi p}{\Omega^{1/3}}, \quad p \in G, \quad G=\left[-\frac{N^{1/3} - 1}{2},\frac{N^{1/3} - 1}{2}\right]^3 \subset \mathbb{Z}^3.
\end{equation}
Additionally, let $G_0 = G$\textbackslash \{(0,0,0)\} be the set of allowed frequencies excluding the singular zero mode.

The terms of the first-quantized electronic structure Hamiltonian can be expanded in the plane wave basis in a way that is commensurate with the linear combinations of unitaries (LCU) approach:
\begin{align}
    T &= \frac{\pi^2}{\Omega^{2/3}} \sum_{j=1}^\eta  \sum_{w\in {x,y,z}}
        \sum_{r=0}^{n_p - 2} \sum_{s=0}^{n_p - 2} 2^{r+s} \sum_{b\in{0,1}} \left( \sum_{p\in G}
        (-1)^{b(p_{w,r}p_{w,s}\oplus 1)} \ket{p}\bra{p}_j\right), \\
    U &= \sum_{\nu\in G_0} \sum_{A=1}^{N_A} \frac{2\pi\zeta_A}{\Omega\Vert k_\nu\Vert^2}
        \sum_{j=1}^\eta \sum_{b\in{0,1}} \left( -e^{-ik_\nu\cdot R_A} \sum_{q\in G}
        (-1)^{b[(q-\nu)\notin G]} \ket{q-\nu}\bra{q}_j\right), \\
    V &= \sum_{\nu\in G_0} \frac{\pi}{\Omega\Vert k_\nu\Vert^2} \sum_{i\neq j=1}^\eta
        \sum_{b\in{0,1}} \left(\sum_{p,q\in G} (-1)^{b[(p+\nu)\notin G]\lor[(q-\nu)\notin G]}
        \ket{p+\nu}\bra{p}_i \cdot \ket{q-\nu}\bra{q}_j\right)
\end{align}
where $n_p = \lceil \log(N^{1/3}+1) \rceil$ is the number of qubits needed to store the one component of the momentum $p_w$, $\eta$ is the number of electrons, $\ket{p}\bra{q}$ is shorthand for $I_1\otimes \cdots\otimes (\ket{p}\bra{q})_j\otimes\cdots\otimes I_\eta$, and $\zeta_A$ and $R_A$ are the atomic charge and position of nuclei $A$, respectively, with $N_A$ total nuclei. The norms of each of these terms is given by
\begin{equation}
    \norm{T} = \frac{6\eta\pi^2}{\Omega^{2/3}}(2^{n_p-1} - 1)^2, \quad
    \norm{U} = \frac{\eta \norm{\zeta}}{\pi\Omega^{1/3}}\norm{\nu}, \quad
    \norm{V} = \frac{\eta (\eta - 1)}{2\pi\Omega^{1/3}}\norm{\nu},
\end{equation}
where
\begin{equation}
    \norm{\zeta} = \sum_{A=1}^{N_A} \zeta_A, \quad
    \norm{\nu} = \sum_{\nu\in G_0} \frac{1}{\Vert \nu \Vert^2},
\end{equation}
and the total norm of the Hamiltonian is given as $\norm{} = \norm{T} + \norm{U} + \norm{V}$.

For simplicity of discussion, we assume $\alpha_l > 0$ and therefore $\prep{} = \unprep{}$ below.
One naive implementation of the block encoding would be to simply construct $\prep{}$ and $\sel{}$ oracles for each term $T, U,$ and $V$ separately;
however, it turns out that preparing the shared momentum state depending on the $\norm{\nu}$ term in $U$ and $V$ is a large source of complexity, which motivates splitting the prepare oracles into a $\prep{T}$ and $\prep{U+V}$.
If we had access to operations such as $\prep{T}, \prep{U+V}, \sel{T},$ and $\sel{U+V}$ with block encodings
\begin{align}
\begin{split}
    \bra{0}\prep{T}^\dagger\sel{T}\prep{T}\ket{0} &= \frac{T}{\norm{T}}, \\
    \bra{0}\prep{U+V}^\dagger\sel{U+V}\prep{U+V}\ket{0} &= \frac{U+V}{\norm{U}+\norm{V}},
\end{split}
\end{align}
then we could combine them into a block encoding for the full Hamiltonian as
\begin{align}
\begin{split}
    \prep{H} &= \left(\sqrt{\frac{\norm{T}}{\norm{}}}\ket{0} +
    \sqrt{\frac{\norm{U}+\norm{V}}{\norm{}}}\ket{1} \right)
    \otimes \prep{T} \otimes \prep{U+V}, \\
    \sel{H} &= \ket{0}\bra{0}\otimes\sel{T}\otimes I + \ket{1}\bra{1}\otimes I\otimes\sel{U+V}.
    \label{eq:prep_sel_T_U+V}
\end{split}
\end{align}

However, the momentum state needed for $\prep{U+V}$ can only be done efficiently with low magnitude of success;
specifically, Berry \textit{et al.} \cite{berry2018improved} provides a procedure to efficiently construct the momentum state that allows us to implement
\begin{equation}
    \prept{U+V} \ket{0}\otimes I \approx \frac{1}{2}\ket{0}\otimes \prep{U+V} 
        + \frac{\sqrt{3}}{2}\ket{1}\otimes\prep{U+V}^\perp.
    \label{eq:prept_U+V}
\end{equation}
We must now either (i) use amplitude amplification (AA) to boost the success probability of applying $\prep{U+V}$ (but with increased cost), or (ii) find a way to use $\prept{U+V}$ directly and account for the new normalization factor of $4(\norm{U}+\norm{V})$.
Su \textit{et al.} \cite{su2021fault} show how to use the latter approach and introduce an additional auxiliary qubit which renormalizes the $T$ and $U+V$ terms in Eq.~\ref{eq:prep_sel_T_U+V}.
Assuming that $\norm{T} < 3(\norm{U}+\norm{V})$ for simplicity, the resulting block encoding is given by
\begin{align}
\begin{split}
    \prep{H} &= \left(\cos(\theta)\ket{0} + \sin(\theta)\ket{1} \right)
        \otimes \prep{T} \otimes \prept{U+V}, \\
    \sel{H} &= \underbrace{I\otimes I\otimes\ket{0}\bra{0}\otimes\sel{U+V}
        }_{\ubmltext{$\prep{U+V}$ succeeded, \\ apply $\sel{U+V}$}} + 
        \underbrace{\ket{0}\bra{0}\otimes\sel{T}\otimes\ket{1}\bra{1}\otimes I
        }_{\ubmltext{$\prep{U+V}$ failed and first qubit 0, \\ apply $\sel{T}$}} +
        \underbrace{\ket{1}\bra{1}\otimes I\otimes\ket{1}\bra{1}\otimes I
        }_{\ubmltext{$\prep{U+V}$ failed and first qubit 1, \\ apply neither}},
    \label{eq:prept_sel_T_U+V}
\end{split}
\end{align}
where $\theta = \arccos{\sqrt{\norm{T}/3(\norm{U}+\norm{V})}}$ is needed in order to rebalance the terms and recover a block encoding with the proper normalization factor $\norm{T}+\norm{U}+\norm{V}$.

\subsubsection{Implementation of \texorpdfstring{$PREP \prep{H}$}{H}}
The full state we want to prepare is
\begin{equation}
\begin{split}
    &\underbrace{\big(\cos(\theta)\ket{0}_a + \sin(\theta)\ket{1}_a\big)
    }_{\ubmltext{Balances T and U+V terms}}
    \quad\underbrace{\ket{+}_b}_{\ubmltext{Indexes bit values\\Used in $T,U,V$}} \\ 
    &\otimes \frac{1}{\eta} \underbrace{\left( \sqrt{\eta - 1}\ket{0}_c\sum_{i\neq j = 1}^\eta
    \ket{i}_d \ket{j}_e + \ket{1}_c\sum_{j=1}^\eta\ket{j}_d \ket{j}_e \right)
    }_{\ubmltext{Register $c$ flags $i\neq j$, $d$ and $e$ index electrons\\
    All used in $V$, just $e$ used in $T,U,V$}}
    \underbrace{\left( \frac{1}{\sqrt{3}}\sum_{w=0}^2\ket{w}_f\right)
    }_{\ubmltext{Indexes $x,y,z$ components\\Used in $T$}}\\ &\otimes
    \underbrace{\left( \frac{1}{2^{n_p-1}-1} \sum_{r,s=0}^{n_p-2} 2^{(r+s)/2} \ket{r}_g\ket{s}_h \right)
    }_{\ubmltext{$r$ and $s$ each index bits in momentum\\Used in $T$}}
    \underbrace{\left( \sqrt{\frac{\norm{U}}{\norm{U}+\norm{V}}}\ket{0}_i +
    \sqrt{\frac{\norm{V}}{\norm{U}+\norm{V}}}\ket{1}_i \right)
    }_{\ubmltext{Balances $U$ and $V$ terms}}\\ &\otimes
    \underbrace{\left( \sqrt{\frac{p_\nu}{\lambda_\nu}}\ket{0}_j \sum_{v\in G_0}
    \frac{1}{\Vert\nu\Vert}\ket{\nu}_k + \sqrt{1-p_\nu}\ket{1}_j\ket{\nu^\perp}_k \right)
    }_{\ubmltext{Register $j$ heralds good state, i.e. flag qubit in Eq.~\ref{eq:prept_U+V}\\
    Register $k$ indexes momenta\\
    Used in $U,V$}}
    \underbrace{\left( \frac{1}{\sqrt{\norm{\zeta}}}
    \sum_{A=1}^{N_A} \sqrt{\zeta_A}\ket{R_A}_l \right)
    }_{\ubmltext{Stores nuclear charge and position\\Used in $U$}},
    \label{eq:prep_tuv_state}
\end{split}
\end{equation}
where the underbraces give a short description, including which Hamiltonian term they are used in, for each part of the state.
Su \textit{et. al.} \cite{su2021fault} assigns these terms in the following way: $\prep{T}$ prepares registers $b,f,g,h$; $\prept{U+V}$ prepares registers $c,d,e,j,k,l$; and the rotations needed for registers $a$ and $i$ are performed separately.
Some of this assignment is arbitrary in that some registers are useful to multiple terms and could be performed by either oracle.
Also note that register $l$ is being used to store the nuclear positions $R_A$ rather than the index $A$, saving one set of auxiliary qubits.

Since explicit circuits have been shown for various components of these oracles in the literature~\cite{lee2021even,su2021fault,delgado2022Apr,babbush2019quantum}, we simply give a high-level overview of the procedure for each register in Eq.~\ref{eq:prep_tuv_state} below.

\begin{enumerate}
    \item Rotate the qubit in register $a$ to select between $T$ and $U+V$
    \item Perform a Hadamard on register $b$ to index the bit values $\{0,1\}$
    \item Prepare equal superpositions of $\eta$ values in registers $d$ and $e$ using the method described in Appendix A.2 of \cite{lee2021even}
    \item Test for equality between registers $d$ and $e$, storing the result in register $c$
    \item Prepare equal superpositions of 3 values in register $f$
    \item Prepare exponential superpositions in registers $g$ and $h$ using the method described in Section II.B of \cite{su2021fault}
    \item Prepare an equal superposition of $n_{\eta\zeta} = \eta + 2\norm{\zeta}$ values and perform an inequality test to determine the weighting between $U$ and $V$ terms in register $i$
    \item Perform the momemtum state preparation procedure from \cite{babbush2019quantum} in register $k$, with register $j$ as the flag qubit
    \item Load nuclear charges $\sqrt{\zeta_A}$ and positions $R_A$ into register $l$ using QROM
\end{enumerate}

While the sizes of many of these registers are well-defined based on their limits in Eq.~\ref{eq:prep_tuv_state}, there are three notable exceptions that we enumerate here.
Register $a$ has size $n_T$, which determines the precision of the rotation by angle $\theta$.
As part of the momentum state preparation, there is a register of size $n_M$ that determines the precision of the inequality test that computes the $1/\Vert\nu\Vert$ coefficient.
Finally, the size of register $l$ is given by size $n_R$, which determines the precision of the nuclear positions $R_A$ that are loaded in from QROM.
The values of $n_T$, $n_M$, and $n_R$ contribute to the error of the block encoding and therefore impact the resource estimates, as shown below.
 
\subsubsection{Implementation of \texorpdfstring{$\sel{H}$}{SEL H}}
The $\sel{}$ oracles should perform the transformations
\begin{equation}
\begin{split}
    \sel{T}: &\ket{b}_b \ket{j}_e \ket{w}_f \ket{r}_g \ket{s}_h \ket{p_j}\\
    &\mapsto (-1)^{b(p_{w,r}p_{w,s}\oplus 1)}\ket{b}_b \ket{j}_e \ket{w}_f \ket{r}_g \ket{s}_h \ket{p_j},\\
    \sel{U}: &\ket{b}_b \ket{j}_e \ket{0}_i \ket{\nu}_k \ket{R_A}_l \ket{q_j}\\
    &\mapsto -e^{-ik_\nu\cdot R_A}(-1)^{b[(q-\nu)\notin G]}\ket{b}_b \ket{j}_e \ket{0}_i \ket{\nu}_k \ket{R_A}_l \ket{q_j - \nu},\\
    \sel{V}: &\ket{b}_b \ket{i}_d \ket{j}_e \ket{1}_i \ket{\nu}_k \ket{p_i} \ket{q_j}\\
    &\mapsto (-1)^{b\left([(p+\nu)\notin G]\lor [(q-\nu)\notin G]\right)} \ket{b}_b \ket{i}_d \ket{j}_e \ket{1}_i \ket{\nu}_k \ket{p_i+\nu} \ket{q_j-\nu}.
    \label{eq:sel_tuv_maps}
\end{split}
\end{equation}
As all three $\sel{}$ operations require performing arithmetic on the momentum registers, this is most efficiently done by first controlling a swap of the momentum into auxiliary registers, performing the operation on the auxiliary, and then swapping back.

As with the prepare oracles, we note that these oracles are shown in detail in the literature~\cite{su2021fault,delgado2022Apr} and summarize the high-level overviews for the oracles in Eq.~\ref{eq:sel_tuv_maps} below.
For $\sel{T}$, the procedure is
\begin{enumerate}
    \item Controlled on register $f$, copy component $w$ of momentum $p$ into an auxiliary register
    \item Controlled on register $g$ ($h$), copy bit $r$ ($s$) of $p_w$ into an auxiliary qubit, respectively
    \item Controlled on the auxiliary qubits containing $p_{w,r}$ and $p_{w,s}$, perform a phase flip on register $b$ (unless both bits are one)
    \item Erase the auxiliary registers (with Cliffords and measurement)
\end{enumerate}
while for $\sel{U}$ and $\sel{V}$, the procedure is
\begin{enumerate}
    \item ($U$ and $V$) Controlled on registers $d$ and $e$, swap momentum registers $p$ and $q$ into auxiliary registers
    \item ($V$ only) Controlled on $\ket{0}$ for ``$V$ only" qubit, add $\nu$ into the auxiliary register for $p$
    \item ($U$ and $V$) Controlled on $\ket{0}$ for ``$U$ or $V$" qubit, subtract $\nu$ into the auxiliary register for $q$
    \item ($U$ and $V$) If $p+\nu$ and $q-\nu$ are inside the box $G$, perform a phase flip on register $b$
    \item ($U$ and $V$) Controlled on registers $d$ and $e$, swap the auxiliary registers back to the momentum registers
    \item ($U$ only) Controlled on $\ket{0}$ for ``$U$ or $V$" qubit and $\ket{1}$ for ``$V$ only" qubit, apply the phase $e^{-ik_\nu\cdot R_A}$
\end{enumerate}

\subsubsection{Logical qubit counts: Hamiltonian}
A full accounting of qubit costs are given in Appendix C.1 of \cite{su2021fault}, but the number of logical qubits needed is
\begin{equation}
\begin{split}
    N_{\rm qubits} & = 3\eta n_p + n_{\eta\zeta} + 2n_\eta + 3n_p^2 + 17n_p + \max(n_T, n_R + 1) + 5n_R + 5n_M + 4n_M n_p + 31
    \label{eq:first_quant_qubits}
\end{split}
\end{equation}

\subsubsection{T counts: Hamiltonian}
The $T$ cost of block encoding the first-quantized electronic structure Hamiltonian is
\begin{equation}
\begin{split}
    T[\mathcal{U}(\hamiltonian)] = \underbrace{7}_{\ubmltext{Toffoli to $T$\\conversion \\from \cite{nielsen2002quantum}}} \Big[&
    \underbrace{2(n_T + n_{\eta\zeta} + 2b_r - 12)}_{
    \ubmltext{Prepare registers $a$ and $i$,\\i.e. rotation between $T$ and $U+V$\\and superposition for $U$ and $V$ test}} \quad+ 
    \underbrace{14n_\eta + 8b_r - 36}_{
    \ubmltext{Prepare registers $c,d$ and $e$,\\i.e. equal superpositions over electron\\indices and $i\neq j$ test}} \\
    & + \underbrace{2(2n_p + 9)}_{
    \ubmltext{Prepare registers $f,g,$ and $h$,\\i.e. equal superpositions over\\momentum components and bits}} +
    \underbrace{12\eta n_p + 4\eta - 8}_{
    \ubmltext{Controlled swaps of \\$p$ and $q$ registers\\for $\sel{}$ operations}} \quad + \quad
    \underbrace{5(n_p - 1) + 2}_{\ubmltext{Cost of $\sel{T}$}}\\
    & +  \underbrace{a\left[ 3 n_p^2 + 15 n_p - 7 + 4 n_M(n_p + 1)\right]}_{
    \ubmltext{Prepare registers $j$ and $k$,\\i.e. momentum state}} + 
    \underbrace{\norm{\zeta} + \text{Er}(\norm{\zeta})}_{
    \ubmltext{Prepare register $l$,\\i.e. $\sqrt{\zeta_l}$ amplitudes\\and QROM for $R_l$}}\\
    & + \underbrace{24n_p}_{\ubmltext{Add/subtract $\nu$ into\\momentum registers}} + 
    \underbrace{8N_pn_R}_{\ubmltext{Phasing by $-e^{ik_\nu\cdot R_A}$}} +
    \underbrace{18}_{\ubmltext{Selection between\\$T,U,$ and $V$}}\Big]
    \label{eq:1quant_hamiltonian_t}
\end{split}
\end{equation}
where $b_r$ is the number of bits of precision for rotation on an auxiliary qubit and can be taken as $b_r=7$ for a success probability of over 0.999, $a$ is a factor that is 1 if no amplitude amplification (AA) for $\prept{U+V}$ is used and 3 if a single round of AA is used, and
\begin{equation}
    \text{Er}(x):= \min_{k\in\mathbb{N}}\left(2^k + \lceil 2^{-k}x\rceil\right)
\end{equation}
is the cost of erasing the output of $x$ entries of QROM.

However, we must consider that we rarely want to apply the block encoded Hamiltonian in isolation;
rather, we want to use controlled applications of the Hamiltonian both in $\mathcal{U_{\rm sp}(\hamiltonian)}$ and $\Lambda(\mathcal{W}(\hamiltonian))$ in Fig. \ref{fig:circuit_diagram}.
Of the qubits in Eqn.~\ref{eq:first_quant_qubits}, there are $n_{\eta\zeta} + 2n_\eta + 8n_p + n_M + 16$ qubits that need to be controlled/reflected on.
Therefore, the full $T$ cost of the controlled Hamiltonian is
\begin{equation}
     T[\Lambda(\mathcal{U}(\hamiltonian))] =  T[\mathcal{U}(\hamiltonian)] + 4\left[n_{\eta\zeta} + 2n_\eta + 8N_p + n_M + 16 \right]
\end{equation}

Finally, we return to the quantities $n_T, n_M,$ and $n_R$, which introduce errors
\begin{equation}
\begin{split}
    \epsilon_T &= \frac{\pi\norm{}}{2^{n_T}},\\
    \epsilon_M &= \frac{2\eta}{2^{n_M}\pi\Omega^{1/3}}(n_{\eta\zeta}
-1)(7\times2^{n_p+1}-9n_p-11-3\times2^{-n_p}),\\
    \overlap_R &= \frac{\eta\norm{\zeta}}{2^{n_R}\Omega^{1/3}}\sum_{\nu\in G_0}\frac{1}{\Vert\nu\Vert}.
    \label{eq:epsilton_TMR}
\end{split}
\end{equation}
We determine $n_T, n_M,$ and $n_R$ by setting their corresponding errors such that $\epsilon_T = \epsilon_M =
\epsilon_R = \epsilon/10$, where $\epsilon$ is the total error in the calculation.
This is done such that the error from these sources is smaller than those corresponding to coming from Theorem 1 or rotation synthesis discussed in Appendix B. 
We then performed a grid search over an array of values, starting 4 below and ending 4 above the calculated values.
We found that the grid search only improved the $T$ counts by one or two percent.
We also compared our values to those found by the TFermion package~\cite{casares2022tfermion}, which performs a gradient-free optimization over these parameters.
Empirically, these analytical $T$ counts were within a factor of 2 of the TFermion counts for several test cases we studied.

\subsection{T counts and qubits for Quantum Phase Estimation (QPE)}
\setcounter{equation}{0}
\setcounter{figure}{0}
\renewcommand{\theequation}{E.4.\arabic{equation}}
\renewcommand{\thefigure}{E.4.\arabic{figure}}
In this work we use the low-T cost QPE algorithm of Babbush \textit{et al.}~\cite{babbush2018encoding}, shown in Fig~\ref{fig:circuit_diagram}(a).
All of the following details can be found in that work.
The circuit is composed of three components, the $qpe$ register initialization $\chi_m$, a sequence of controlled Szegedy walk operators $\Lambda(\mathcal{W}(\mathcal{H}))$, and an inverse Fourier Transform $QFT^\dagger$ to read out the final measured energy.
The expense of implementing the phase estimation algorithm is determined entirely by two different parameters: $\Delta E$, the target Holevo variance of the measured ground state energy, and $\alpha \geq || \mathcal{H} ||$.
The agregate quantity $p$, defined below in terms of $\Delta E$ and $\alpha$, will be used to parameterize the cost of QPE:
\begin{equation}
    p = \lceil \log(\frac{\sqrt{2}\pi\alpha}{2\Delta E}) \rceil.
\end{equation}

\subsubsection{T counts}
We break up the T counts into the three different circuit components. 
Beginning with $\chi_m$, the only non-Clifford gates required are $p$ controlled -single-qubit rotation operators, which would require four single-qubit rotation operators each and a two CNOT gates to implement.
For $QFT^\dagger$, we require $p(p-1)/2$ controlled-single-qubit rotations, which again each require four single-qubit rotations. 
We note also that the rotation synthesis error in $\chi_m$ and $QFT^\dagger$ required to ensure a Holevo variance $\Delta E$ is $\epsilon_{QFT}/(\pi p)$ where $\epsilon_{QFT} \leq 2^{-(p+1)}$.
For the Szegedy walk, we require $2^p$ applications of $\mathcal{H}$, with the T counts for the latter denoted $T[\mathcal{U}({\mathcal{H}})]$, and can be located in Eq~\ref{eq:tfim_hamiltonian_t} and Eq~\ref{eq:1quant_hamiltonian_t}.
The total T count for QPE then amounts to:
\begin{equation}
    T[QPE] = (4p + 2p(p-1)) (10 + 4\log_2 \frac{\pi p}{\epsilon_{QFT}}) + 2^p T[\mathcal{U}({\mathcal{H}})]
\end{equation}

\subsubsection{Logical qubits}
Only $p$ qubits are required for QPE to measure the energy with Holevo variance $\Delta E$.

\renewcommand{\theequation}{E\arabic{equation}}
\renewcommand{\thefigure}{E\arabic{figure}}
\setcounter{figure}{0}

\end{document}